\documentclass[final,5p,longtitle]{elsarticle}

\usepackage{etoolbox}

\makeatletter
\patchcmd{\MaketitleBox}
  {\hrule\vskip12pt}
  {\newpage\hrule\vskip12pt}
  {}{}
\makeatother

\pdfoutput=1
\usepackage{hyperref,amsmath,amssymb,amsfonts,gensymb,comment,stackengine,graphicx,xcolor,lmodern,mathtools,siunitx,subcaption,textcomp}
\usepackage[T1]{fontenc}
\usepackage{tabularray,multirow,lineno}
\usepackage{tikz}
 
\hypersetup{
    colorlinks=true,
    linkcolor=blue,
    urlcolor=cyan,
    citecolor=green,
    filecolor=magenta
}
\journal{NIM A}
\newcommand{\til}{{\raise.17ex\hbox{$\scriptstyle\sim$}}}

\newcommand{\dbar}{\ensuremath{\overline{\textrm{d}}}}

\newcommand{\pbar}{\ensuremath{\overline{\textrm{p}}} }
\newcommand{\pbarNS}{\ensuremath{\overline{\textrm{p}}}}
\newcommand{\hebar}{\ensuremath{\overline{^3\textrm{He}}} }

\newcommand{\hebarNS}{\ensuremath{\overline{^3\textrm{He}}}}

\newcommand{\quant}[2]{\ensuremath{{#1}\,\mathrm{#2}}}
\newcommand{\squant}[2]{\ensuremath{\til{#1}\,\mathrm{#2}}}

\hypersetup{
  pdftitle  = {The General Antiparticle Spectrometer (GAPS) Antarctic Balloon Payload},
  pdfauthor = {Field Rogers and Gabriel Bridges},
  pdfkeywords = {Balloon instrumentation; Scintillation detectors; Silicon detectors; Passive temperature control; Cosmic ray; Dark matter detector}
}

\begin{document}
\newpageafter{title}
\newpageafter{author}
\newpageafter{abstract}

\begin{frontmatter}
    \title{The General Antiparticle Spectrometer (GAPS) Antarctic Balloon Payload}
    
    \author{The GAPS Collaboration}
    \author[JAXA]{Kazutaka Aoyama}
    \author[NEU]{Tsuguo Aramaki}
    \author[UCLA]{Padrick Beggs}
    \author[Trieste,IFPU]{Mirko Boezio}
    \author[UCSD]{Steven E.\ Boggs}
    \author[Trieste]{Valter Bonvicini} 
    \author[CU]{Gabriel Bridges\corref{cor1}}
    \ead{glb2139@columbia.edu}
    \author[Napoli]{Donatella Campana}
    \author[SSL]{Scott Candey}
    \author[SSL]{William W.\ Craig}
    \author[UH]{Philip von Doetinchem}
    \author[NEU]{Conor Earley} 
    \author[UCLA]{Erik Everson}
    \author[ORNL]{Lorenzo Fabris}
    \author[UCLA]{Sydney Feldman}
    \author[JAXA]{Hideyuki Fuke}
    \author[CU]{Florian Gahbauer}
    \author[UH]{Cory Gerrity}
    \author[Pavia,Bergamo]{Luca Ghislotti}
    \author[CU]{Charles J.\ Hailey}
    \author[UCLA]{Takeru Hayashi}
    \author[Tokai]{Akiko Kawachi}
    \author[Tokai]{Kai Konoma}
    \author[ROIS]{Masayoshi Kozai} 
    \author[Pavia,Bergamo]{Paolo Lazzaroni}
    \author[SSL]{Alexander Lowell}
    \author[Pavia,Bergamo]{Massimo Manghisoni} 
    \author[Rome]{Matteo Martucci}
    \author[Tohoku]{Keita Mizukoshi}
    \author[Trieste]{Emiliano Mocchiutti} 
    \author[SSL]{Brent Mochizuki}
    \author[Shinshu]{Kazuoki Munakata}
    \author[Trieste,IFPU]{Riccardo Munini}
    \author[JAXA]{Shun Okazaki}
    \author[Heliospace]{Jerome Olson}
    \author[UCLA]{Rene A.\ Ong}
    \author[Napoli]{Giuseppe Osteria}
    \author[Rome]{Francesco Palma} 
    \author[CU]{Kaliro\"e Pappas}
    \author[CU]{Kerstin Perez}
    \author[Napoli,UNapoli]{Francesco Perfetto}
    \author[Pavia,UPavia]{Lodovico Ratti}
    \author[Pavia,Bergamo]{Valerio Re}
    \author[Pavia,Bergamo]{Elisa Riceputi}
    \author[Chicago]{Brandon Roach}
    \author[SSL]{Field R.\ Rogers\corref{cor1}}
    \ead{fieldr@berkeley.edu}
    \author[FNAL]{Nathan Saffold}
    \author[Kanagawa]{Suzuto Sakamoto} 
    \author[Rome,UTrento]{Pratiksha Sawant}
    \author[Napoli,UNapoli]{Valentina Scotti}
    \author[Kanagawa]{Yuki Shimizu}
    \author[Rome,URome]{Roberta Sparvoli}
    \author[UH]{Achim Stoessl} 
    \author[NEU]{Arathi Suraj}
    \author[Firenze,UFirenze]{Alessio Tiberio}
    \author[UH]{Grace Tytus}
    \author[Firenze]{Elena Vannuccini}
    \author[CU]{Sarah Vickers}
    \author[Rome,URome2]{Luigi Volpicelli}
    \author[SJTU]{Zhen Wu}
    \author[SJTU]{Mengjiao Xiao}
    \author[SJTU]{Jinghe Yang}
    \author[CU,MIT]{Kelsey Yee}
    \author[JAXA]{Tetsuya Yoshida}
    \author[Trieste]{Gianluigi Zampa}
    \author[NEU]{Jiancheng Zeng}
    \author[UCLA]{Jeffrey Zweerink}
    \cortext[cor1]{Corresponding authors}
    
    \address[JAXA]{Institute of Space and Astronautical Science, Japan Aerospace Exploration Agency (ISAS/JAXA), Sagamihara, Kanagawa 252-5210, Japan.}
    \address[NEU]{Northeastern University, 360 Huntington Avenue, Boston, MA 02115, USA.}
    \address[UCLA]{University of California, Los Angeles, Los
    Angeles, CA 90095, USA.}
    \address[Trieste]{INFN, Sezione di Trieste, I-34149 Trieste, Italy.}
    \address[IFPU]{IFPU, I-34014 Trieste, Italy.}
    \address[UCSD]{Department of Astronomy \& Astrophysics, University of California, San Diego, La Jolla, CA 90037, USA.}
    \address[CU]{Columbia Astrophysics Laboratory, Columbia University, New York, NY 10027, USA.}
    \address[Napoli]{INFN, Sezione di Napoli, I-80126 Naples, Italy.}
    \address[SSL]{Space Sciences Laboratory, University of California, Berkeley, 7 Gauss Way, Berkeley, CA 94720, USA.}
    \address[UH]{University of Hawaii at Manoa, Honolulu, HI 96822 USA.}
    \address[ORNL]{Oak Ridge National Laboratory, Oak Ridge, TN 37831, USA.}
    \address[Pavia]{INFN, Sezione di Pavia, I-27100 Pavia, Italy.}
    \address[Bergamo]{Università di Bergamo, I-24044 Dalmine (BG), Italy.}
    \address[Tokai]{Tokai University, Hiratsuka, Kanagawa 259-1292, Japan. }
    \address[ROIS]{Polar Environment Data Science Center, Co-creation Center for Future Science Systems, Research Organization of Information and Systems, Tachikawa, Tokyo 190-0014, Japan.} 
    \address[Rome]{INFN, Sezione di Rome "Tor Vergata", I-00133 Rome, Italy.}
    \address[Tohoku]{Tohoku University, Aoba-6-3 Aramaki, Aoba Ward, Sendai, Miyagi 980-8578, Japan.}
    \address[Shinshu]{Shinshu University, Matsumoto, Nagano 390-8621, Japan.}
    \address[Heliospace]{Heliospace Corporation, 2448 6th St., Berkeley, CA 94710, USA.}
    \address[UNapoli]{Università di Napoli "Federico II", I-80138 Naples, Italy. }
    \address[UPavia]{Università di Pavia, I-27100 Pavia, Italy.}
    \address[Chicago]{Kavli Institute for Cosmological Physics, University of Chicago, 5640 S. Ellis Ave., Chicago, IL 60637, USA.}
    \address[FNAL]{Fermi National Accelator Laboratory, Batavia, IL 60510, USA.}
    \address[Kanagawa]{Kanagawa University, Yokohama, Kanagawa 221-8686, Japan. }
    \address[UTrento]{Università di Trento, Dipartimento di Fisica, V. Sommarive 14, Trento, Italy.}
    \address[URome]{Università di Roma ``Tor Vergata'', I-00133 Rome, Italy. }
    \address[Firenze]{INFN, Sezione di Firenze, I-50019 Sesto Fiorentino, Florence, Italy.}
    \address[UFirenze]{Università degli Studi Firenze, 50121 Firenze FI, Italy.}
    \address[URome2]{Università di Roma "La Sapienza", Dipartimento di Fisica, P. le A. Moro 2, Rome, Italy.}
    \address[SJTU]{State Key Laboratory of Dark Matter Physics, Key Laboratory for Particle Astrophysics and Cosmology (MoE), Shanghai Key Laboratory for Particle Physics and Cosmology, School of Physics and Astronomy, Shanghai Jiao Tong University, 800 Dongchuan Rd. Minhang District, Shanghai 200240, China.}
    \address[MIT]{Massachusetts Institute of Technology, Cambridge, MA 02139, USA.}

    \begin{abstract}
    The General Antiparticle Spectrometer (GAPS) is an Antarctic stratospheric balloon mission designed to provide unmatched sensitivity to low-energy (\quant{<0.25}{GeV/{\it n}}) cosmic-ray antiprotons, antideuterons, and antihelium nuclei as signatures of dark matter. 
    The distinctive GAPS particle identification technique relies on measuring the energy loss along the track of an incoming antinucleus as it slows down and is captured into an exotic atom, and then detecting the de-excitation X-rays and the nuclear annihilation products.
    This measurement is realized using a Tracker composed of more than 1000 custom silicon strip detectors and a
    plastic scintillator time-of-flight (TOF) system instrumenting more than \quant{40}{m^2}. Together, these subsystems provide the velocity and energy resolution, stopping power, particle tracking, and X-ray identification necessary to distinguish rare antinucleus signals from the abundant positive-nucleus backgrounds, all within the constraints of a high-altitude mission. 
    A multi-loop capillary heat pipe system has been developed to maintain the tracker operating temperature with significant mass and power savings over a conventional pump-based system. 
    The first GAPS science payload flew for 25 days during the 2025/26 NASA Antarctic balloon campaign. 
    We detail the design, integration, and commissioning of the payload prior to flight. 
    \end{abstract}


\end{frontmatter}

\section{Introduction}\label{sec:intro}

The General Antiparticle Spectrometer (GAPS) \cite{Mori,HaileyGAPS1,HaileyGAPS2} is the first experiment optimized to detect low-energy {(kinetic energy $\lesssim$\quant{0.25}{GeV/{\it n}})} cosmic-ray antideuterons (\dbar). 
The discovery of cosmic-ray \dbar\ by GAPS would be a smoking-gun signature of new physics such as dark matter (DM) \cite{DonatoAntideuterons,Donato08,Donato17,DbarReview, AntideuteronAramakiSensitivity}. 
GAPS will also measure the cosmic-ray antiproton (\pbarNS) spectrum in a lower {energy regime} than any other experiment \cite{AntiprotonAramakiSensitivity,GAPSAntiprotonSensitivity} and {deliver} unprecedented sensitivity to low-energy antihelium-3 nuclei (\hebarNS) \cite{GapsHeBarSensitivity}. 
In addition to antinuclei, GAPS will also produce spectra for low-energy cosmic-ray positive nuclei including protons, deuterons, and helium-3. 
The GAPS 
sensitivity is based on several multi-week stratospheric flights on NASA's Antarctic long-duration balloon (LDB) platform. 

The unique strength of low-energy cosmic-ray \dbar\ as a DM signal lies in the ultra-low astrophysical \dbar\ background. Astrophysical \dbar\ production is strongly kinematically suppressed --- especially at energies below a few GeV/{\it n}
 --- and is predicted to be several orders of magnitude below the signal expected from realistic DM models \cite{DbarReview}. 
The GAPS Antarctic flight program will provide {nearly} two orders-of-magnitude improvement in \dbar\ sensitivity comparee to the current limits \cite{AntideuteronAramakiSensitivity,BessDbar24}, extending into an unprobed energy range where the significant DM signals would exist. 

GAPS uses a novel particle identification technique based on the formation, de-excitation, and annihilation of antinucleonic exotic atoms, which provides robust rejection of positive nuclei and discrimination between antinucleus species. 
In 2004, this detection technique was demonstrated by the GAPS Collaboration in an antiproton beam at the KEK accelerator, with measurement of de-excitation and decay products from antiprotonic exotic atoms in agreement with modeling predictions \cite{KEK06,KEK13}. 
In 2012, a prototype GAPS balloon payload demonstrated critical hardware performance, including the silicon technology at the heart of the GAPS detection scheme, from a high-altitude platform \cite{DoetinchemPGAPS,MognetPGAPS,FukePGAPS,Fuke17}. 
Elements of the thermal system, including a radiator operationg below $-50\degree$C,  were demonstrated in two additional test flights \cite{Okazaki24}. 

In this paper, we describe the payload of the first Antarctic flight of the GAPS program. 

\autoref{sec:design} introduces the design considerations required to achieve DM sensitivity through a cosmic-ray \dbar\ measurement. \autoref{sec:engineering} discusses the engineering solutions that meet these requirements, including the mechanical, thermal, electronics, and data/ communication systems. The two sensitive detector systems, the silicon Tracker and the plastic scintillator time-of-flight (TOF) System, are detailed in \autoref{sec:trk} and \autoref{sec:tof}, respectively. {The full system readout is in \autoref{sec:sys}.}

\section{GAPS Design Principles}\label{sec:design}

\begin{figure*}[t]
\centering
\includegraphics[width=\textwidth,trim = 0 0 0 0,clip]{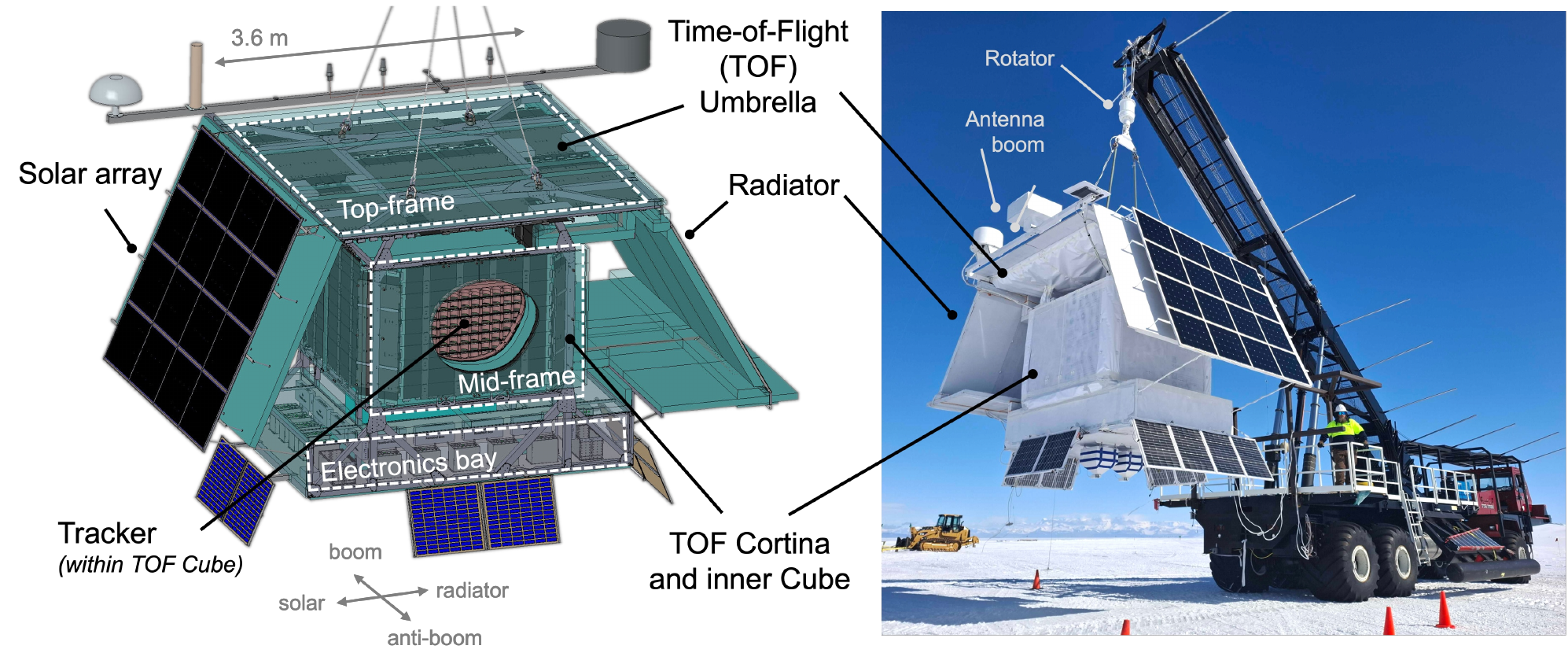}
\caption{\label{fig:gaps-inst}
GAPS design ({\it left}) and photo of the completed instrument ({\it right}). 
GAPS consists of a Tracker with $>1000$ custom Si(Li) detectors, surrounded on all sides by a plastic scintillator TOF system. 
Supporting electronics and on-board computing are housed in the lower electronics bay. An antenna boom supported by the Top-frame supports antennas used for telemetry. 
A MCHP thermal system transports heat from the Tracker volume to a radiator.
Power is provided by an array of sixteen solar panels. 
The payload is suspended by four cables from the flight train, which includes a rotator that keeps the panels oriented toward the Sun and the radiator oriented toward space. 
As illustrated, gondola coordinates are defined according to the solar side opposite the radiator side and the boom side opposite the `anti-boom' side.}
\end{figure*}

Identification of low-energy cosmic-ray antinuclei requires: 
\begin{enumerate}
    \item a high-altitude polar platform, 
    as these particles are strongly deflected to the poles by the Earth's magnetic field and attenuated by the atmosphere,
    \item a combination of large instrument acceptance and long observation time for sensitivity to the rare \dbar\ fluxes of the order $10^{-6}$\,[m$^{2}$\,s\,sr\,(GeV/{\it n})]$^{-1}$ predicted from dark matter, and
    \item rejection of the positive nucleus backgrounds, which exceed the predicted \dbar\ signals by a factor of $>10^9$ in the energy band of interest, as well as robust discrimination between \pbar\ and rare \dbar\ or \hebar\ signals.
\end{enumerate}

When an antinucleus in the GAPS energy range interacts with matter, it loses energy via ionization and excitation, slowing down until its kinetic energy is comparable to its binding energy with the target material. 
Then, an atom in the target material captures the antinucleus with probability of order unity, forming an exotic atom in an excited state. The exotic atom de-excites, with the lower-level atomic transitions producing X-rays, followed by nuclear capture and annihilation to hadrons (primarily pions) \cite{KEK06,KEK13}. 
Each antinucleus species can be distinguished based on the combination of its incident energy-loss patterns and the resulting X-ray energies and hadron multiplicity. 
The abundant positive nuclei do not form exotic atoms and can thus be rejected. 
Compared to magnetic spectrometer experiments, such as BESS~\cite{BessPolarPbar}, PAMELA \cite{Picozza07}, and AMS-02~\cite{Aguilar21}, this technique offers larger instrument acceptance, due to increased area and mass that can be dedicated to sensitive detectors; a lower energy range, due to decreased inactive material; and improved identification power for these rare-event signatures.

The GAPS instrument is illustrated in \autoref{fig:gaps-inst}. 
The Tracker is a central \quant{1.6}{m}$\times$\quant{1.6}{m}$\times$\quant{1.0}{m} volume with 10 layers. For the first flight, the top seven layers are populated with custom lithium-drifted silicon (Si(Li)) sensors read out via a custom Application-Specific Integrated Circuit (ASIC). The bottom three layers are populated with aluminum disks (analogous to the Si(Li) sensors) that contribute to the target mass and resistive heaters used for thermal control. In total, the Tracker provides a target capable of stopping \dbar\ with energies up to \quant{0.25}{GeV}/{\it n}, measuring energy depositions from \quant{20}{keV} to \quant{100}{MeV}, and providing approximately cm-scale resolution for particle tracking, sufficient for the scientific goals. 
The TOF provides a high-speed system trigger, particle velocity and energy deposition measurements, and further timing and position information for particle tracking. 
The inner TOF Cube surrounds the Tracker with >98\% hermeticity, with an outer Umbrella positioned \squant{90}{cm} from the top of the Cube and an outer Cortina positioned \quant{>30}{cm} from the sides of the Cube.
The Tracker is cooled to an operating temperature of $<-35\degree$C by a multi-loop capillary heat pipe (MCHP) system that transports heat from the warm Tracker electronics to a \squant{9}{m^2} radiator oriented towards space. Sixteen solar panels provide \quant{1.3}{kW} of power, and all supporting electronics and on-board computing are housed in the lower electronics bay.

\begin{figure*}[tb]
\centering
\includegraphics[width=0.9\textwidth]{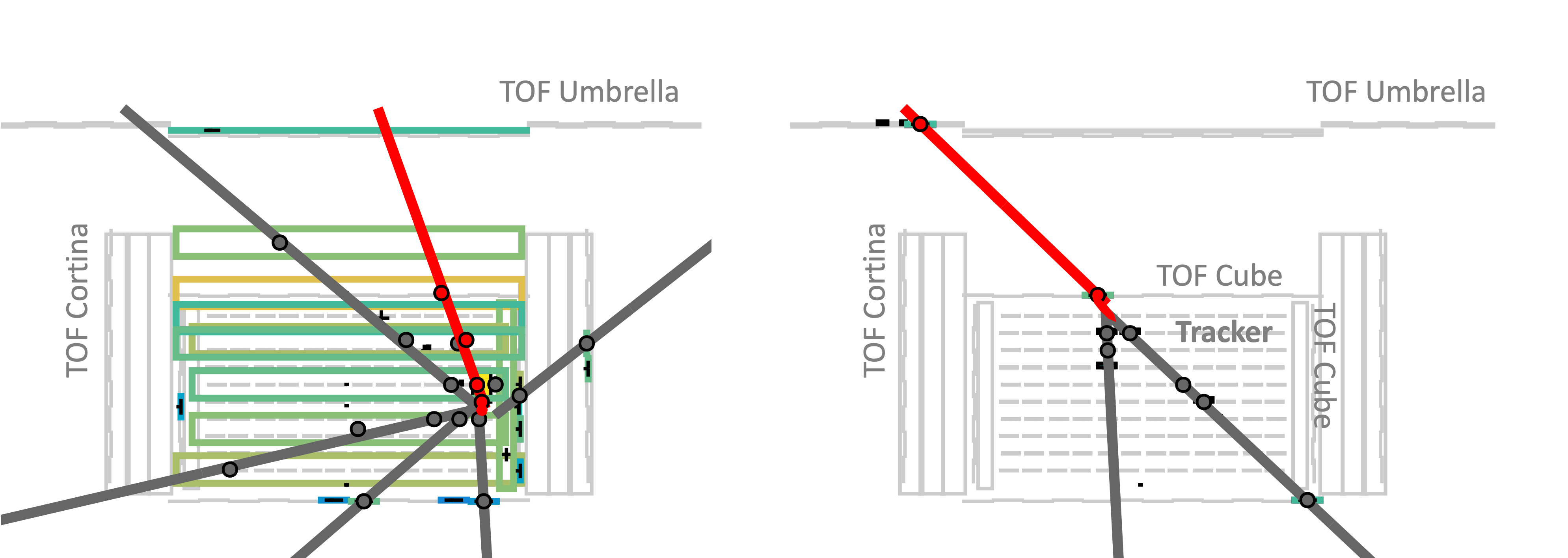}
\caption{\label{fig:pid}
 Reconstructed tracks and vertex in a simulated antideuteron event ({\it left}) and a candidate cosmic-ray interaction event from ground commissioning data ({\it right}). Lines illustrate reconstructed tracks, and energy depositions in active sensor elements are indicated by the color scale. The GAPS simulation and reconstruction framework contains the full instrument mass model, although only active detector elements are illustrated.}
\end{figure*}

\autoref{fig:pid} illustrates the characteristic event topology of a low-energy \dbar\ in simulations and a candidate interacting event from ground commissioning data. 
The simulated \dbar\ first traverses two layers of the TOF, which measures the incident velocity and energy loss rates. 
The particle slows down and is captured into an exotic atom in the Tracker, which measures the remaining d$E$/d$x$ losses, de-excitation X-rays (not pictured), and the charged particle tracks of the nuclear annihilation products. The TOF additionally records at least one energy deposition for each charged particle track produced in the annihilation. Together, this information facilitates reconstruction of the common annihilation vertex and the primary and secondary particle tracks~\cite{Reconstruction}. 

NASA's Antarctic LDB (\quant{0.97}{million\,m^3}, 34H\footnote{\href{https://www.csbf.nasa.gov/balloons.html}{https://www.csbf.nasa.gov/balloons.html}}) platform provides a unique capability for weeks-long exposures (typical flight lengths of $18-21$ days, with a maximum recorded flight duration of 57.3 days \cite{Miles25}) from a high altitude (up to \squant{37}{km}) in a polar region. 
The balloon platform and the stratospheric environment introduce an interconnected set of engineering constraints on mass, geometry, mechanical robustness, thermal, power, and remote command and readout. \autoref{sec:engineering} details how these constraints have been addressed.


\section{Engineering Design}\label{sec:engineering}
\subsection{Gondola Mechanical Design}\label{sec:mech}

The gondola frame is divided into three substructures: the Top-frame, Mid-frame, and Electronics bay (E-bay), as illustrated in \autoref{fig:gaps-inst}. This frame must withstand shocks up to \squant{8}{g} vertical and \squant{4}{g} horizontal on the total instrument weight of \quant{2875}{kg}. It is primarily composed of \quant{7.62}{cm}-wide \quant{0.32}{cm}-thick 6061-T6 Al square tubing, selected for its strength to weight ratio.

The Top-frame measures \squant{370}{cm}$\times$\quant{370}{cm} and supports the TOF Umbrella, antenna boom, solar array, and radiator. The flight-train links to the corners of a \squant{1.5}{m}$\times$\quant{2.2}{m} load-bearing rectangle in the center of the Top-frame. A \quant{4.5}{kg} control unit\footnote{This unit is responsible for un-sealing the balloon after inflation is complete.} must be jettisoned from the balloon during launch and could impact the payload. The top-frame is reinforced with a cross-hatched layer of polyester straps which prevent this control unit from damaging the scintillators in the Umbrella. A continuous load path runs from the four flight-train links through the Top-frame, the four Mid-frame vertical members, and down to the E-bay.

The Mid-frame is the central \squant{229}{cm}$\times$\quant{225}{cm}$\times$\quant{170}{cm} structure which supports the Tracker, TOF Cube, and TOF Cortina.
The Tracker and TOF Cube form an inner \squant{164}{cm}$\times$\quant{165}{cm}$\times$\quant{137}{cm} unit. To minimize thermal conduction between the cold Tracker and the gondola frame, this inner structure is connected to the Mid-frame and E-bay with four Delrin\textsuperscript{\textregistered} blocks. 

The E-bay is the \squant{371}{cm}$\times$\quant{225}{cm}$\times$\quant{61}{cm} base of the gondola and contains all GAPS flight electronics, except for nine of the TOF readout and trigger units (see \autoref{sec:tofrat}), as well as the NASA/CSBF Support Instrumentation Package (SIP)\footnote{\href{https://www.nasa.gov/wp-content/uploads/2025/06/ldb-support-for-science-el-100-10-h-rev-b.pdf}{https://www.nasa.gov/wp-content/uploads/2025/06/ldb-support-for-science-el-100-10-h-rev-b.pdf}}. Four NASA/CSBF solar arrays, which provide power to the SIP, and two ballast hoppers are mounted below the E-bay.

\subsection{Thermal Design}\label{sec:thermal}

\subsubsection{Tracker Cooling System}\label{sec:ohp}

\begin{figure*}[tbh]
\centering
\includegraphics[width=0.9\textwidth]{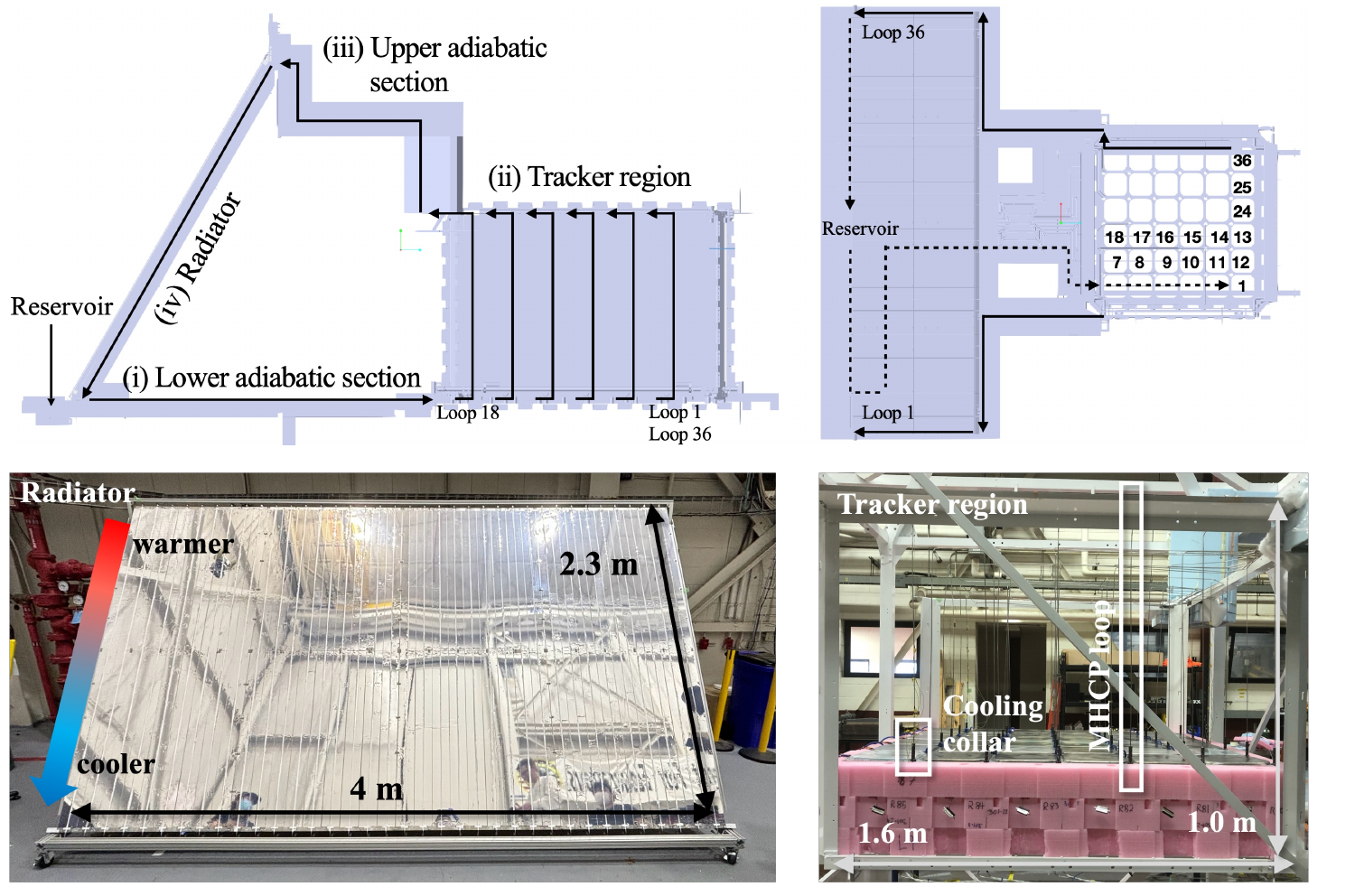}
\caption{\label{fig:tracker-thermal}
Diagram of the MHCP thermal system ({\it top}). 36 MHCP loops transport heat to the radiator ({\it bottom-left}) from the Tracker volume ({\it bottom-right}, shown with first three layers integrated). 
}
\end{figure*}

The Tracker Si(Li) detectors operate at a nominal temperature of \quant{<-35}{\degree C}. A novel MCHP thermal system was developed to transport, with minimal mass, up to \quant{300}{W} of heat (\quant{120}{W} produced by Tracker electronics and \quant{180}{W} of environmental load) over horizontal and vertical distances of $\gtrsim$\quant{1}{m} away from the Tracker to a radiator oriented towards space. In principle, the system can operate indefinitely and with minimal power.
Basic functions of this unique system have been verified by environmental testing, a series of test flights, and numerical analyses \cite{Okazaki24,OHP,CapillaryHeatPipe,FukeOHPIEEE,Fuke24}.
    
The MCHP is a hybrid of an oscillating heat pipe (OHP, or pulsating heat pipe) and a thermosyphon. It consists of capillary tubes filled with a two-phase fluid that flows via passive circulation (self-oscillation): expanding into vapor in the heated section and condensing into liquid in the cooled section. The capillary tubes are designed to maintain a homogeneous flow so that locally liquid and vapor move together. Each MCHP loop comprises (i) a lower adiabatic section, (ii) a heating section, (iii) an upper adiabatic section, and (iv) a cooling section. The heating and cooling sections in each loop have identical lengths.
Circulation of the working fluid is driven by the density difference between the vapor evaporated in the Tracker volume and the liquid condensed in the radiator. Multiple loops connected in series excite a self-oscillation, which further drives circulation. Both of these drivers are entirely passive.
    
GAPS uses 36 MCHP loops connected in series, which enter the Tracker as a $6\times6$ array, as illustrated in \autoref{fig:tracker-thermal}. Each loop is made from a 3\,mm outer diameter, 1.3\,mm inner diameter stainless steel tube. Every other loop is equipped with a check valve to ensure directional flow. The radiator is composed of a \quant{3.0}{mm} thick, \quant{2.3}{m}$\times$\quant{4.0}{m} 6061-T6 aluminum alloy plate, with each loop bonded to a groove in the plate with a methacrylate adhesive. The radiator is coated with a $0.12$\,mm layer of silver-coated Teflon fluorinated ethylene propylene (FEP), which has low solar absorptance and a high infrared (IR) emissivity (absorptance of $0.09$ and emissivity of $0.75$). The working fluid is trifluoromethane (R23), chosen for its low boiling point ($-82\degree$C) and its ability to maintain a two-phase state over a wide temperature range.

The majority of the R23 in the system is contained in a liquid reservoir which is attached to the first loop of the MCHP. The temperature (and, by extension, the saturation pressure) of the MCHP can be controlled with a dedicated heater attached to this reservoir. Additionally, an auxiliary heater is attached to each loop immediately beneath the Tracker, which can be used to induce boiling of the R23 should the system become too cold for self-oscillation to begin. Temperature monitoring and heater control of the MCHP system were implemented using custom-designed hardware and software, including Field-Programmable Gate Array (FPGA) logic, multichannel Analog-to-Digital Converter (ADC) boards, and a power control unit.

\begin{figure*}[th]
\centering
\includegraphics[width=0.9\textwidth]{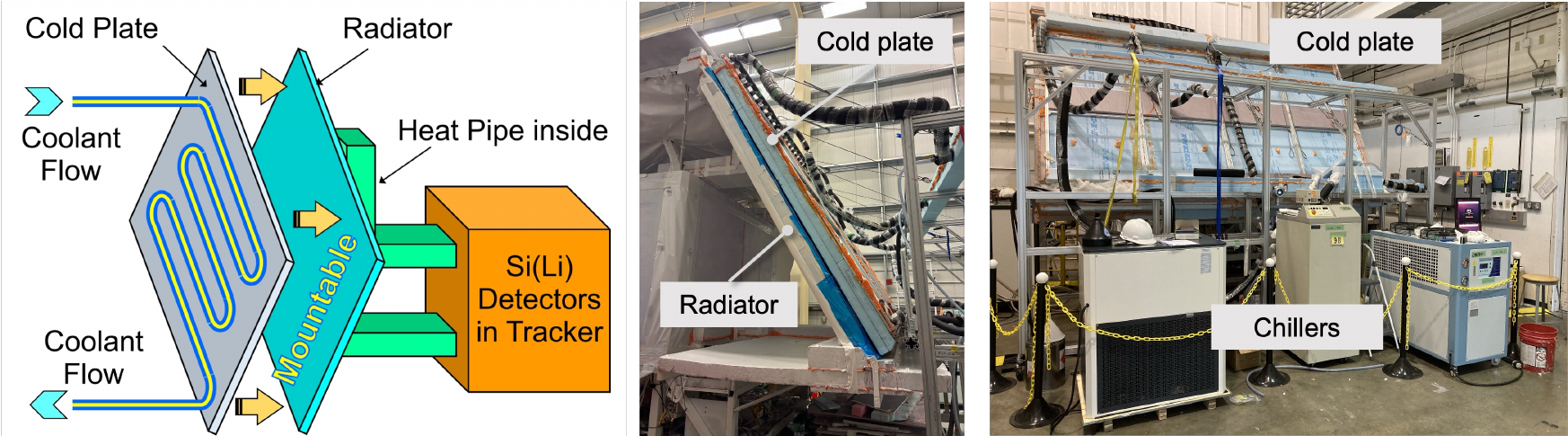}
\caption{\label{fig:GCS}
Diagram of the ground cooling system (GCS, {\it left}) \cite{Fuke23}. An aluminum cold plate is thermally coupled to the radiator ({\it center}) and cooled via a system of commercial chillers ({\it right}).
}
\end{figure*}

As illustrated in \autoref{fig:GCS}, a ground cooling system (GCS) \cite{Fuke23} has also been developed to cool the radiator and thus enable ground calibration of the Tracker. The GCS has a cold plate which is attached to the radiator and is actively cooled using recirculating chillers. Once the cold plate sufficiently cools the radiator, the heat pipes transfer heat from the detectors to the radiator passively, bringing the detectors down to their operating temperature.

\subsubsection{System Thermal Management}\label{sec:therm_balance}

Careful thermal management is necessary to prevent overheating of active or sun-facing elements or overcooling of space-facing elements due to radiative loss.
    
The solar zenith angle during flight is $\til60\degree$, such that the sun-side of the payload and the Top-frame are illuminated. A layer of \quant{4}{in}-thick 
Owens Corning FOAMULAR\textsuperscript{\textregistered} 250 extruded polystyrene foam (hereafter ``foam'') is painted white \footnote{The foam was painted with Rust-Oleum 1990502 Painter’s Touch Latex Paint in Flat White.} and mounted on top of the Top-frame to insulate and reflect solar radiation. An additional foam solar-shield is mounted behind the solar array to fully shade the gondola. 

The radiator must be shielded from both Earth's albedo and gondola radiation. Shielding is achieved using foam sheets mounted to the back of the radiator and spanning from the base of the radiator to the E-bay. The MCHP tubes which transit between the radiator base and Tracker are embedded in this foam sheet, which also insulates those tubes. 

The Tracker is embedded within a foam cube \quant{15.24}{cm} wider than the Tracker on all sides. It is thermally isolated from the TOF Cube by a layer of aluminized Mylar and further insulated from the external environment by a \quant{4}{in}-thick foam layer surrounding the TOF Cube. This outer insulation layer is sealed with expanding polyurethane foam to eliminate convection, which can add significant heat to the MCHP even at float altitude. All mechanical contact between the TOF Cube/Tracker assembly and the gondola is made with non-conductive Delrin\textsuperscript{\textregistered} blocks.

The TOF Umbrella and Cortina are thermally shielded with a mixture of painted foam and reflective white Mylar, which prevents the black TOF panels from overheating. 

Both insulation and radiative surfaces are used around the E-bay 
to maintain the overall temperature of each component within the required range of $+30\degree$C to $-30\degree$C. Heat from the E-bay electronics is radiated away primarily through \quant{0.508}{mm}-thick, white-painted Al skins installed on the radiator, boom, and anti-boom sides of the E-bay. The sun-side and bottom of the E-bay are insulated with \quant{1}{in}-thick foam sheets. Additionally, critical electronics are equipped with heaters for active thermal control should temperatures become too low. Foam on top of the E-bay prevents the electronics from radiating onto the Tracker.

The E-bay electronics units, which are a mixture of custom and off-the-shelf units, are designed with vacuum thermal performance in mind. Off-the-shelf components are industrial grade with temperature rating for most parts spanning at least $-40\degree$C to $+70\degree$C. 
Thermally conductive pads are placed between the units and the gondola frame to ensure adequate heat flow, and most are painted white to improve radiative heat dissipation. 
Prior to integration in the gondola, units were tested in vacuum at the extreme temperatures expected in flight. 

Any gaps in the insulation surrounding the gondola are sealed with a combination of white mylar and white 
duct tape.

\subsection{Gondola Power System}\label{sec:power}

\begin{table}
\begin{centering}
\caption{\label{tab:power_budget} Power budget of the GAPS Payload.}
\begin{tblr}{colspec={|ll|},hlines,width=\textwidth}
\centering
{System}  & {Budget [W]}  \\
\hline
Tracker & 250 \\
TOF & 500\\
Thermal & 250\\
Gondola Control & 100\\
Starlink\textsuperscript{*} & 100 \\ 
\hline
Total & 1200 \\
\end{tblr}\newline
\small
\textit{\raggedright *Apart from Starlink (discussed in \autoref{sec:tmcm}), the units responsible for telemetry and commands are powered with the SIP and are not included in this budget.}
\end{centering}
\end{table}

Sixteen \quant{100}{W} photovoltaic (PV) panels from SunCat Solar LLS allow substantial margin on the \squant{1}{kW} power budget given in \autoref{tab:power_budget}, after accounting for transmission losses and temperature de-rating at float altitude. This budget does not include the SIP and its associated electronics, which were operate on an independent power system.
The panels are held in a 4$\times$4 array within a {\quant{2.76}{m}$\times$\quant{3.19}{m}} frame held at an angle of $23\degree$ from vertical for optimal irradiation in Antarctica. 

Eight Odyssey PC1100 
\quant{12}{V} lead-acid batteries are connected pairwise for a \quant{24}{V} system. Considering voltage drop in transmission, the payload receives power at \quant{19-29}{V} depending on the battery charge state. 
Each battery delivers \squant{42}{Ah} in ideal temperature conditions, which translates to \squant{4}{kWh} for the system. Battery capacity is most important during ascent, before gondola pointing is established. 
Charging is controlled by two {Morningstar TriStar MPPT 60} solar battery chargers, 
each connecting eight PV panels to the battery bank. 

Battery power is distributed to the gondola electronics using four custom 8-channel power distribution units (PDUs). Redundant 
RS-485 serial interfaces on each PDU allow separate control of each channel. Continuous battery voltage is also provided to a NASA/CSBF-provided science stack unit, which enables control of six critical PDU channels, including the onboard computer and network used to control the gondola, when the control computer itself is powered off.

\subsection{Gondola Control System}\label{sec:gcu}
The gondola control system manages the data acquisition activities of the Tracker and TOF subsystems, facilitates packaging and telemetry of both science and housekeeping data, and provides both automated and adjustable controls to ensure ongoing health of the instrument during remote operation. It consists of the Gondola Control Unit (GCU; the central computer), an auxiliary GCU, two Ethernet switches, a serial server, a Raspberry Pi unit, and a Labjack unit. 

The GCU features a Versalogic EBX-38 CPU, \quant{4}{GB} Versalogic VL-MM9-4EBN RAM, redundant \quant{1}{TB} Swissbit solid-state drives, three Ethernet ports with distinct internet protocol addresses, and direct serial connections to the PDUs, serial server, and SIP, all powered by a custom power board. 
The CPU is thermally coupled to an Al block bolted directly to the E-bay floor to prevent overheating. The auxillary GCU uses the same hardware but contains \quant{4}{TB} ADATA solid-state drives for additional on-board storage. 

The Raspberry Pi and Labjack are custom units built around the respective commercial parts. The Labjack aids in maintaining safe operating temperatures for all of the gondola electronics by reading in temperatures from LM135 sensors and controlling Kapton heaters. The Raspberry Pi unit interfaces additional heaters and temperature sensors within the battery bank and reads out other battery housekeeping. 

\subsection{Telemetry and Commands}\label{sec:tmcm}
As summarized in \autoref{tab:tmcm}, GAPS uses five distinct systems for telemetry downlink and command uplink, all managed by NASA/CSBF. The downlink is almost entirely based on {the User Datagram Protocol (UDP)}, though two links also allow for remote access to the GCU via Transmission Control Protocol (TCP). 

These links each have distinct advantages and disadvantages: two support TCP, their bandwidth spans six orders of magnitude, and they have different levels of Antarctic flight heritage and reliability. Iridium Short Burst has the best geographical coverage but features very limited bandwidth; conversely, the GAPS flight serves as a testbed for Starlink on an Antarctic balloon platform, with the possibility for significantly higher downlink speeds. 
Together, the links allow remote configuration of the instrument in flight and downlink of housekeeping and science data.

\begin{table}
\caption{\label{tab:tmcm} Summary of the telemetry and commands pathways. The speeds indicated are for the telemetry downlink.}
\centering
\vspace{1mm}
\begin{tblr}{colspec={|lll|},hlines,width=\textwidth}
{System}  & {Bandwidth} & {Protocol}  \\
\hline
Line-of-sight (LOS)$^*$ & \quant{12}{Mbps} & UDP\\
Iridium Short Burst& \quant{15}{kb} / \quant{15}{min} & UDP   \\ 
{Tracking and Data \\ Relay Satellite \\ System (TDRSS)} & \quant{\le300}{kbps} & UDP \\ 
Iridium Pilot & \quant{\le120}{kbps} & UDP, TCP\\ 
Starlink & \quant{4+}{Mbps} & UDP, TCP \\ 
\end{tblr}
\newline
\small
\textit{\raggedright *The line-of-sight (LOS) link is operational while in range of the radio-frequency transmitter and receiver, typically for the first day of flight. The bandwidth enables intensive initial commissioning and calibrations.}\\
\end{table}

\section{Tracker}\label{sec:trk}

\begin{figure*}[tbh]
\centering
\includegraphics[width=0.9\textwidth]{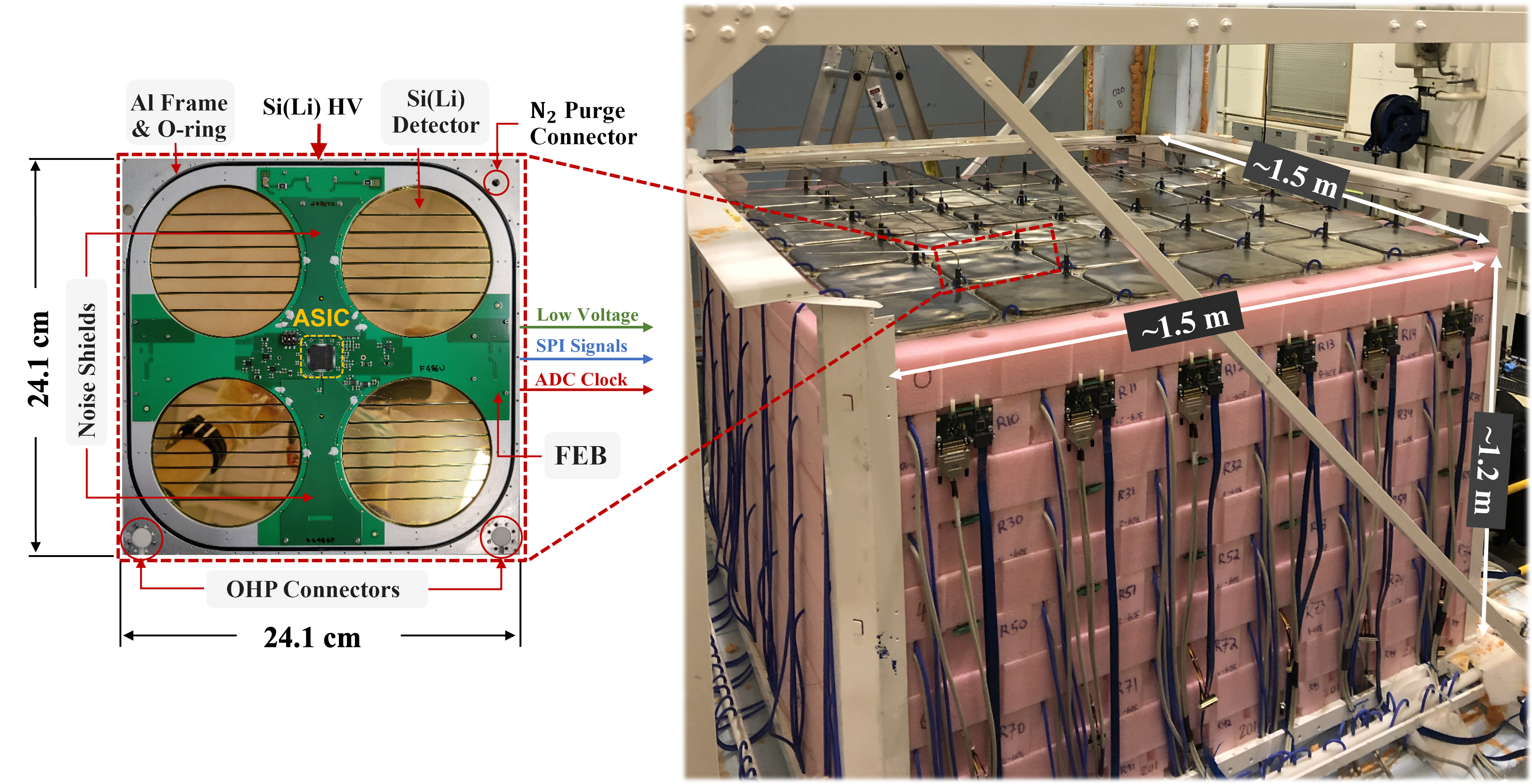}
\caption{\label{fig:tracker}
A Tracker module ({\it left}; top aluminized polypropylene window removed) contains four Si(Li) detectors. The Tracker structure ({\it right}) is pictured after integration of the final layer of 36 modules.}
\end{figure*}

\begin{figure*}[tbh]
\centering
\includegraphics[width=0.9\textwidth]{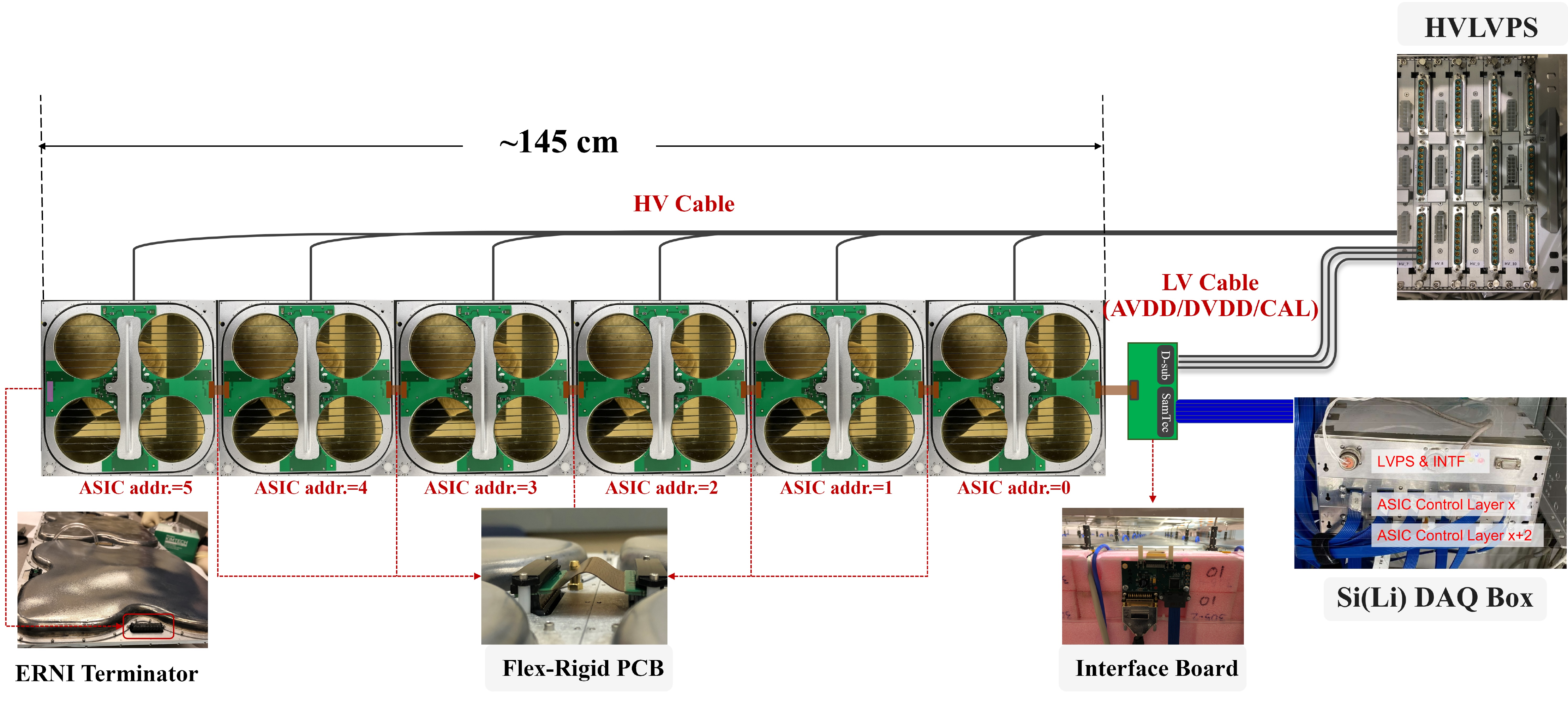}
\caption{\label{fig:tracker_row}
A tracker row illustrated with power and readout connections.}
\end{figure*}

The GAPS Tracker (\autoref{fig:tracker}) provides the stopping power and precision energy measurements required for a low-energy antideuteron search. It provides:

\begin{enumerate}
    \item the high-acceptance ($\gtrsim 15\,\mathrm{m}^2\cdot\mathrm{sr}$) required to meet the sensitivity goals of GAPS given the finite exposure of an LDB flight,
    \item the target mass required to stop cosmic-ray antiparticles with energies up to \quant{0.25}{GeV/}{\it n},
    \item centimeter-scale position resolution required to reconstruct multi-track events,
    \item $\lesssim10\%$ energy resolution up to \squant{100}{MeV} for tracking slow-moving charged particles, and
    \item $<5\,$keV (FWHM) energy resolution between \quant{20}{keV} and \quant{100}{keV} for resolving exotic-atom X-rays.
\end{enumerate}

The tracker meets these requirements using custom lithium-drifted silicon (Si(Li)) detectors and ASIC readout electronics. The detector design is described in \autoref{sec:sili}. Mechanical integration of the Tracker is presented in \autoref{sec:tkr_mech}, and \autoref{sec:tkr_electronics} describes the power distribution and readout systems, including the ASIC architecture, required for Tracker operation.

\subsection{Custom Si(Li) Detectors}\label{sec:sili}

A custom Si(Li) detector fabrication method was developed to satisfy the GAPS
requirements. The large-area detector fabrication techniques, originally demonstrated in-house~\cite{PerezSiLi} were further developed in partnership with Shimadzu Corp, a commercial producer of Si(Li) devices who offered large-scale facilities and trained technical staff for mass production. The detectors were characterized with a $^{109}$Cd source and it was found that $\sim90\%$ of fabricated strips achieved the necessary X-ray energy resolution for exotic atom spectroscopy\cite{RogersSiLi, Xiao23}.

Each GAPS Si(Li) detector is \quant{2.5}{mm} thick and has a diameter of \quant{10}{cm}. The outermost \quant{2}{mm} are machined to a thickness of \quant{1}{mm}, forming a ``top hat'' geometry, for lithium containment during drifting. A \quant{2}{mm} wide guard ring surrounds the active area of the detector, isolating the signal readout from large surface currents. The central \quant{9}{cm}-diameter active area is segmented into eight equal-area strips. Each detector is protected from environmental contaminants by a thin layer of polyimide adhered to the silicon using a passivation procedure developed in-house \cite{SaffoldSiLi}. These detectors operate at a nominal \quant{250}{V} reverse bias and temperatures $\lesssim-30\degree$C.

\subsection{Tracker Mechanical Integration}\label{sec:tkr_mech}

\autoref{fig:tracker} {\it (left)} illustrates a single detector module containing four Si(Li) detectors and the front-end readout board (FEB). The detectors are mounted to the \quant{0.32}{cm}-thick Al module frame with a combination of fluorosilicone O-rings and G10 clamps which absorb shocks during launch and parachute deployment so the the detectors can be recovered post-flight. Before flight, the detectors are protected from environmental contamination by \quant{0.38}{mm}-thick aluminized polypropylene windows which are sealed to both sides of the module frame. The aluminization of these windows creates a Faraday cage around each module. The guard ring of each detector is directly connected to the Al frame which is, itself, connected to the analog ground of the Tracker power system \cite{Xiao23}. 

The FEB supplies the bias voltage to the detectors and connects each strip to the custom readout ASIC. Each detector strip is linked to the FEB with \quant{25}{\si{\micro\meter}}-thick Al wires. Each FEB is equipped with two ERNI 50-pin MicroSpeed BlindMate SMD connectors (used for signal transmission and low voltage power distribution) and one HiRose connector (used for high voltage power). The details of the tracker readout electronics are discussed further in \autoref{sec:asic}.

The module frames are connected to the MCHP tubes via a ``cooling collar,'' a 
\quant{1.17}{cm}-thick anodized Al clamp. These collars provide excellent thermal contact while maintaining electrical isolation between the MCHP and the Tracker analog ground.

Each layer of the Tracker consists of six module rows, with each row containing six detector modules. The modules within a row are electronically linked with Kapton flex-rigid {printed circuit boards} (PCBs) that transmit signals and low-voltage power through the FEB ERNI connectors. The first module in each row is also connected to an interface board, which connects the row to the Tracker power system and backend data acquisition (DAQ) unit. The last module in each row is fitted with an ERNI connector hosting \quant{100}{\Omega} resistors acting as terminations for the differential signal lines. The internal volumes of the six modules in a row are linked via a series of fluorosilicone tubes which connect to manifolds outside of the Tracker volume. \autoref{fig:tracker_row} illustrates a complete Tracker row including its connections to the power system and backend electronics. 

The analog grounds of every row are linked together in the Tracker power system, so adjacent rows must be electrically isolated. This isolation is achieved by mechanically connecting rows through anodized Al nut plates at the corners of the module frames and separating rows with a layer of Kapton tape. The nut plates are also designed to facilitate recovery of the Si(Li) detectors post-flight. A thin groove, carved into the nut plate surface, allows them to be broken by hand so that entire rows can be expediently removed from the Tracker.

The active Tracker layers are arranged such that the detector strips in each layer are oriented perpendicularly to the layers above and below it. This orientation positions the interface boards on the Sun and anti-boom side of the payload.

\subsection{Tracker Electronics}\label{sec:tkr_electronics}

\subsubsection{Tracker Power System}\label{sec:tkr_pwr}

Power is supplied to the Tracker by a custom combination high-voltage/low-voltage power system (HVLVPS). The two HVLVPS units each contain ten high-voltage cards, ten low-voltage cards, and one Ethernet/control card and are capable of powering five Tracker layers. Each pair of high-voltage/low-voltage cards provides power to three rows. 

The high-voltage cards have 18 channels each (one channel per module) and can supply up to \quant{300}{V} and \quant{6}{\si{\micro A}} per channel. High-voltage channels are bundled into groups of six so that each row can be powered with one cable.

Each row requires four distinct low-voltage power lines: two \quant{2.8}{V} lines (one analog, one digital) for the ASICs and two \quant{3.3}{V} lines, one of which powers the interface board while the other drives the calibration Digital-to-Analog Converter (DAC). Each low-voltage card supplies these four power lines to each of three rows.  

\subsubsection{Tracker Readout Electronics}\label{sec:asic}

Each detector module is read out using a custom 32-channel ASIC \cite{Manghisoni23}. Basic detector signal processing is done by a low-noise charge-sensitive amplifier (CSA) followed by a shaper, a sample-and-hold circuit and a discriminator circuit with an adjustable threshold \cite{Manghisoni21}. The analog readout channel implements dynamic signal compression to achieve the required energy resolution in the X-ray energy range ($\lesssim100\,$ keV) while having the required dynamic range to track charged particles \cite{Manghisoni15, Manghisoni18}.The ASIC digitizes the signals from the analog channels with a single, multiplexed, 11-bit Analog-to-Digital Converter (ADC).

The ASICs are highly configurable, which allows for performance optimization and a variety of operation modes. The ASIC reference current and peaking time used by the shaper are set to \squant{5}{\si{\micro A}} (adjustable by 3-bit DAC) and \squant{1.13}{\si{\micro s}} (one of eight selectable times), respectively. The readout threshold is set with a combination of a coarse 8-bit threshold on each ASIC and a fine 3-bit threshold on each individual analog channel. The threshold is set to one of two values, to permit or prevent X-ray energies from triggering, based on the noise characteristics of a module. The ASIC is also equipped with a 32-bit mask for disabling noisy channels.

To facilitate calibration, each ASIC has several settings for operation. The ASICs can be toggled between two trigger modes, internal or external; between reading out all channels or only those over threshold (zero-suppression); and between reading out the FEB temperature or FEB leakage current. A 16-bit DAC, mounted to the FEB, allows the injection of charge directly into the CSAs of every channel. The injection circuit is also equipped with a configurable mask.

The data acquisition and operation of the ASICs is handled by the backend DAQ. The backend DAQ system is composed of four DAQ card-cage units and the interface boards connected to each row. Each backend DAQ unit contains one Ethernet control card and two (or one) ASIC control cards. The Ethernet control cards receive power from the PDUs, transmit data to and receive commands from the GCU via an Ethernet link, receive external digital \texttt{TRIGGER} and \texttt{EVENT\_ID} signals, and transmit \texttt{BUSY} signals. Each ASIC control card configures and reads out the ASICs on one full Tracker layer using a 6-channel 
FPGA. Each FPGA channel controls one row of the Tracker, and the backend DAQ can control individual ASICs by referencing a 3-bit physical address stored on each FEB.

\begin{figure*}[tbh]
\centering
\includegraphics[width=0.9\textwidth,trim = 180 120 38 23,clip]{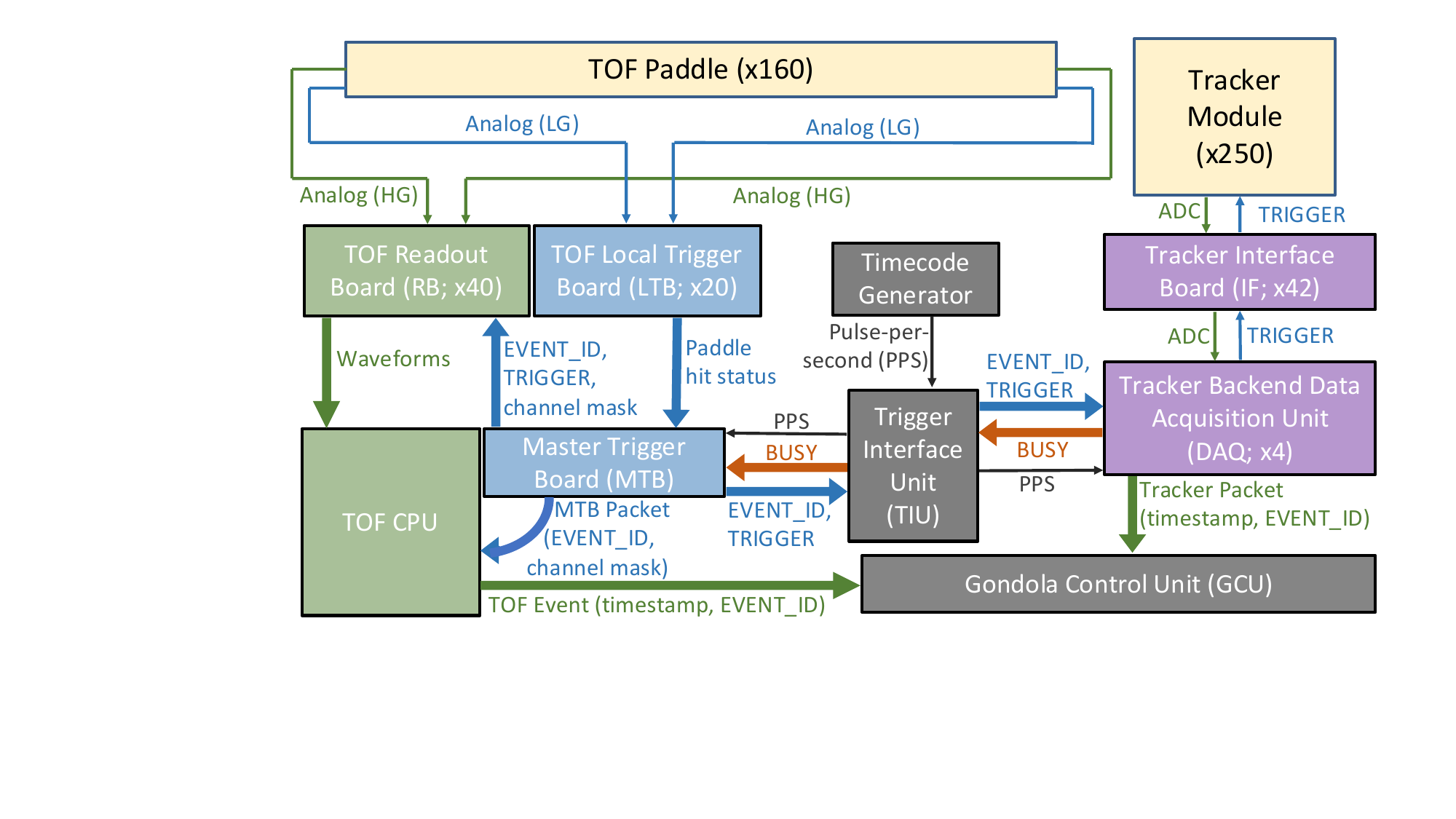}
\caption{\label{fig:electronics}
Overview of the GAPS trigger (blue), busy (orange), and data readout pathways (green). The Tracker electronics, which handle multiple signals on the same board, are illustrated in purple. The \texttt{TRIGGER} and an associated \texttt{EVENT\_ID} are generated by the Master Trigger Board (MTB) based on paddle hit status information from all 20 LTBs processing analog low-gain (LG) signal from eight TOF paddles each. The \texttt{TRIGGER} is distributed directly to the TOF readout boards (RBs) directly and to the Tracker backend DAQ units via the Trigger Interface Unit (TIU). In response to a \texttt{TRIGGER}, data are read out and packaged both from the TOF paddles (high-gain signal; HG) via the RBs and the TOF CPU and from the Tracker modules via the interface boards and the backend DAQ units. The backend DAQs also send a \texttt{BUSY} signal back to the MTB via the TIU, and the TIU distributes a pulse-per-second timestamp to both Tracker and TOF. The Gondola Control Unit manages the merging of events based on matching \texttt{EVENT\_ID}. Timestamps can be used as a backup to the \texttt{EVENT\_ID} system.}
\end{figure*}

As diagrammed in \autoref{fig:electronics}, the Tracker readout process begins when an external \texttt{TRIGGER} signal is received by the backend DAQ. Each backend DAQ unit immediately transmits a \texttt{BUSY} signal to prevent additional triggering during Tracker readout. 
To account for the time difference between the peak of an ASIC signal and the generation of a trigger, the DAQ respects a configurable delay time (set at \quant{800}{ns} during normal operation) before transmitting \texttt{TRIGGER} to the ASICs.
Then, the DAQ requests readout from each ASIC and stores the results in an internal buffer. After readout of the ASICs is complete, the DAQ continues to transmit \texttt{BUSY} for \quant{50}{\si{\micro\second}}  to ensure all ASICs have settled. 
In parallel, the DAQ waits \quant{50}{\si{\micro\second}} from receiving \texttt{TRIGGER} to receive an \texttt{EVENT\_ID} from the TOF, which is used to tag ASIC data prior to transmission to the GCU. If no \texttt{EVENT\_ID} is received in this time, the ASIC data is flagged and assigned a sequential \texttt{EVENT\_ID} by the backend DAQ. 

The deadtime is driven entirely by the Tracker readout and depends strongly on the multiplicity of hits within the Tracker for any given event. The backend DAQ measures the total deadtime every \squant{1}{ms} by recording the cumulative duration of \texttt{BUSY} from all four DAQ units.

\subsection{Tracker Calibration and Operation}\label{sec:tkr_nominal}

\subsubsection{Calibration}

The configurable nature of the ASIC enables detailed calibrations of the tracker. The full set of ASIC calibrations include a pedestal, transfer function, and waveform scan. The pedestal and transfer function are temperature dependent (variable by a few percent for $\Delta T\til5\degree\,$C) and so are routinely measured during flight.

The pedestal, the signal read out when no energy is deposited, is measured by externally triggering an ASIC while zero-suppression is turned off. The transfer function is measured by injecting calibration pulses with a range of amplitudes into the ASIC and recording the response. The precise conversion between deposited energy and digital response can be modeled based on measured properties of the detectors and the front-end electronics and validated by measuring the well-understood energy depositions of minimum-ionizing particles (MIPs).

The waveform scan is performed only once, on the ground, to determine the optimal shaping time for each ASIC. The waveform scan sweeps through the eight shaping times and records a pedestal at each one. The optimal shaping time is the one which produces a pedestal spectrum with the minimum width (for all ASICs the shaping time is \quant{\tau \til 1.13}{\si{\micro\second}}).

The DAQ \texttt{TRIGGER} delay time is also only calibrated on the ground. MIP signals are recorded for a range of delay times, the optimal time is the one which maximized the amplitude of the MIP signal. 

\begin{figure}[tb]
\centering
\includegraphics[width=0.9\columnwidth,trim = 10 10 10 0,clip]{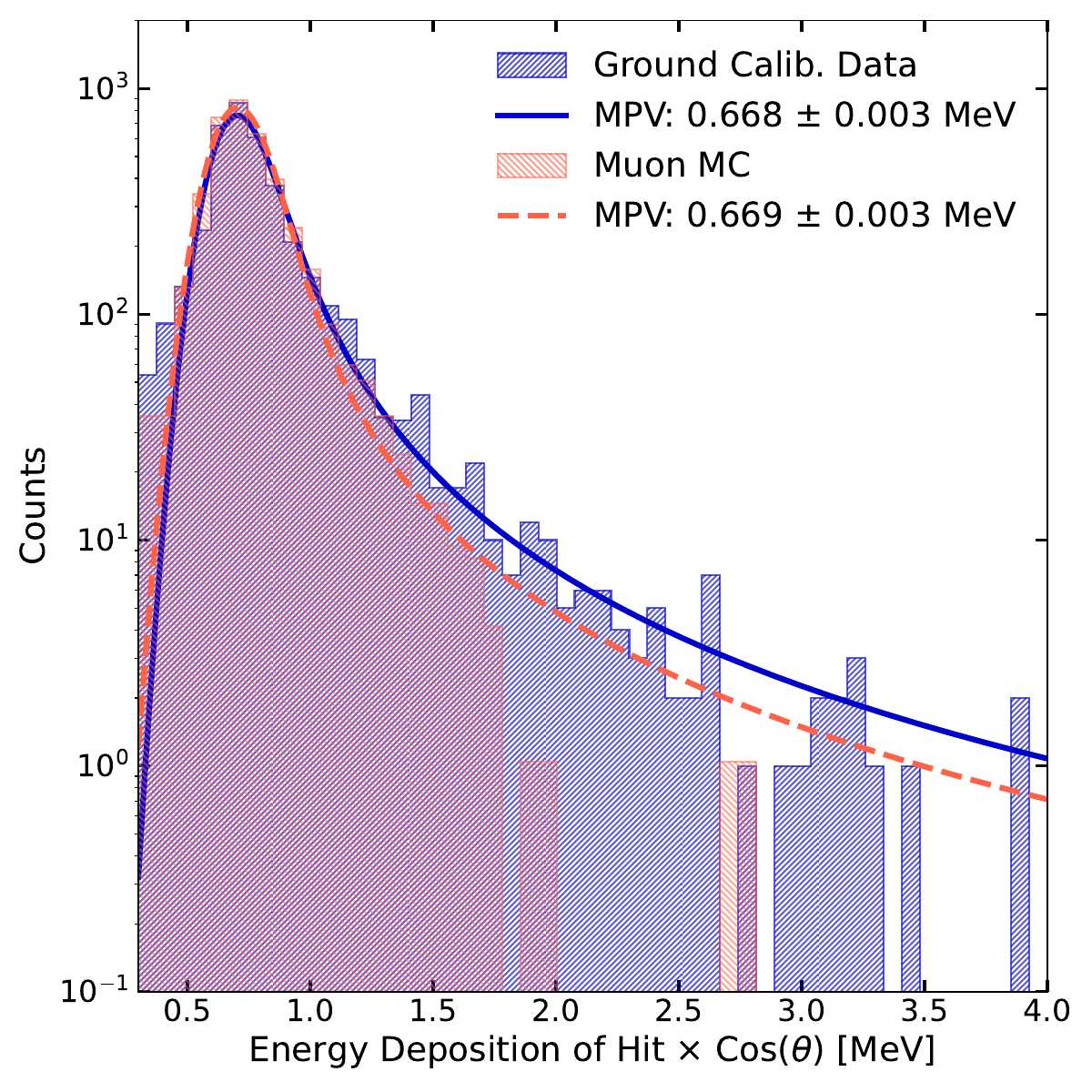}
\caption{\label{fig:tracker_energy}
Energy depositions (corrected for incident angle) from hits on track in a single detector during ground commissioning with cosmic-ray muons, fit to a convolution of a Gaussian distribution with a Landau distribution (blue). The measured distribution is consistent with the distribution expected from simulations of cosmic muon interactions with the instrument (red). The simulated distribution has been normalized to contain the same number of events as were recorded by the detector. 
}
\end{figure}

Performance of the integrated tracker was validated on the ground using a combination of cosmic muons and X-ray sources. Prior to full integration, testing of complete rows with flight-representative power and backend DAQ systems validated that the full readout chain can achieve energy resolution consistent with that obtained at the detector level during calibration \cite{Xiao23}, a result subsequently verified for a subset of detectors after tracker integration. During extensive ground operation, cosmic-ray signals were recorded using an external trigger provided by the TOF. \autoref{fig:tracker_energy} shows agreement between the measured path-length-corrected energy depositions and simulated cosmic muon signals.

\subsubsection{Operation}

During nominal operation, the ASICs are configured to use an external trigger and use zero-suppression to prevent the backend DAQ's buffers from overflowing. Most ASICs are configured with a low energy threshold sufficient for registering exotic atom X-rays. Modules whose noise routinely exceeds this threshold are, instead, configured with a higher energy threshold that permits only MIP-scale depositions. Additional noisy channels are evaluated and have their discriminator masked on a channel-by-channel basis. 

The noise observed in the GAPS Tracker is a combination of intrinsic electronic noise and ``common mode noise'' (CMN) which is common to pairs of detectors. The CMN can be measured and subtracted on an event-by-event basis by reading-out signal-free channels. In practice, the CMN is measured per event by injecting a small calibration pulse into two channels on an ASIC so that those channels are always over threshold.

\section{TOF and System Trigger}\label{sec:tof}

The Time-of-Flight (TOF) system delivers: 
\begin{enumerate}
    \item a system trigger generated within \quant{400}{ns} of a particle interaction, to facilitate readout of the entire instrument within the signal formation and buffer timescales of both sensor systems, 
    \item a measurement of the primary particle's velocity $v$ with $\Delta \beta \equiv \Delta v/c~ \til~ 0.1$ (1$\sigma$) over the full $\beta$ range and $\Delta\beta \til 0.05$ in the signal region, enabling $\beta$-dependent particle identification as well as an energy scale suitable for the antiproton spectral measurement, and
    \item tracking and d$E/$d$x$ information, with hermeticity providing $\ge1$ hit for each charged particle entering or exiting the Tracker to facilitate reconstruction of the primary and secondary tracks. 
\end{enumerate}

Meeting these goals requires resolution of \quant{<400}{ps} on the time-of-flight measurement, position resolution \quant{<6}{cm}, and $>98\%$ hermiticity around the Tracker volume. 
Due to their position on the outside of the payload, some elements of the TOF system experience some of the greatest temperature extremes on GAPS, so care has been taken to ensure mechanical and thermal robustness. 

The plastic scintillator detectors are described in \autoref{sec:pad}. 
Mechanical integration of the scintillator panels is detailed in \autoref{sec:panel}. 
The electronics scheme consists of a central computer and master trigger unit working in tandem with 20 distributed readout and trigger (RAT) units, which are described in \autoref{sec:tofrat}. \autoref{sec:trigscheme} details the trigger generation pathway, \autoref{sec:rb} describes the signal digitization in the RATs, and \autoref{sec:tofevtbuild} describes assembly of paddle-level data into TOF-level events.

\begin{figure*}[h]
\centering
\includegraphics[width=0.9\textwidth]{TOFpp_v3}
\caption{A single TOF scintillator paddle with each end read out by six SiPMs ({\it top}) and a TOF panel with ten paddles ({\it bottom}), assembled with carbon fiber sandwich panels, carbon fiber clamping straps, and polyester strapping.
}
\label{fig:TOF_paddle_panel}
\end{figure*}

\subsection{Scintillators and SiPM Readout}\label{sec:pad}

The 160 scintillator paddles (\autoref{fig:TOF_paddle_panel}) of the TOF are made of polyvinyl toluene (PVT; Eljin Technology part number EJ-200; refractive index n = 1.58). 
All but six are \quant{16}{cm} wide, \quant{0.635}{cm} thick, and \quant{108-180}{cm} long, as specified in \autoref{tab:1pp}.
Each paddle is wrapped with an inner layer of Al foil and an outer layer of vinyl blackout material to achieve internal reflection and light-tightness, respectively. A venting tube inserted into a small opening at the center of the paddle prevents damage to the wrapping during pressure changes. 

The scintillation light is sampled at each paddle end by six silicon photomultipliers (SiPMs; Hamamatsu S13360-6050VE) coupled to each end of the scintillator via \quant{1}{mm}-thick optical silicone cookies.  Each SiPM has an effective photosensitive area of \quant{36}{mm^2}
(14336 pixels). 
The SiPMs at each paddle end are mounted on a preamplifier board that incorporates a transimpendence amplifier to combine the low-impedance outputs of the SiPMs and produce signals capable of driving a \quant{50}{\Omega} cable. The amplifier produces two signal outputs: low-gain (LG, typical MIP pulse height \squant{12.5}{mV} (\autoref{tab:ltbdisc}), for local trigger decisions) and high-gain (HG, typical MIP pulse height \squant{43}{mV}, for the waveform digitizer).
Both the HG and LG outputs are carried to the downstream electronics via Times Microwave Systems LMR-LW \quant{50}{\Omega} coaxial cables.
The SiPMs are operated at a nominal bias of \quant{+58.0}{V}, though adjustments up to \quant{\pm2.0}{V} can be made to correct the collected charge (see \autoref{sec:tofcal}). 

\subsection{TOF Mechanical Integration}\label{sec:panel}

\begin{table*}
\caption{\label{tab:1pp} Summary of the paddles and panels in the TOF system. The TOF system utilizes five distinct panel geometries, with 3, 6, 8, 10, or 12 paddles per panel. The 21 panels and four standalone paddles form the Cube, Cortina, and Umbrella structures of the TOF.}
\begin{centering}
\begin{tblr}{colspec={|llllll|},hlines,width=\textwidth}
{TOF Structure}  & {Component} & N panels&{N paddles per panel} & {Paddle length [cm]} & {Panel area [m$^2$]}\\
\hline
Cube& {sides\\corners\\top/bottom} &{4\\4*\\2}& {8\\1*\\12} & {156 (30), 151 (2){\textsuperscript{$\dagger$}}\\108\\180} & {1.663 (2), 1.635 (2)\\0.108\\3.240}\\
Cortina & {sides\\corners}&{4\\4} & {10\\3} & {172\\151} & {2.546\\0.693} \\ 
Umbrella & {edges\\center}&{6\\1}& {6\\12} &{180 (34), 149 (2){\textsuperscript{$\ddagger$}}\\180} & {1.564 (4), 1.526 (2)\\3.240} \\ 
\end{tblr}
\end{centering}
\newline
\small
\textit{ \raggedright *Denotes the four 10\,cm-wide paddles that are not integrated into panels.}\\
\textit{\raggedright {\textsuperscript{$\dagger$}}The lowest paddles on the Sun and radiator sides of the cube are shorter than the other paddles and only 14.5\,cm-wide, to accommodate the Tracker support structures below and behind these panels.}\\
\textit{\raggedright {\textsuperscript{$\ddagger$}}In two of the 6-paddle panels of the Umbrella, a 149\,cm-long paddle replaces one of the 180\,cm-long paddles, creating an opening through which the diagonal support connecting the Top-frame to the Mid-frame passes. For panel types with multiple form factors, the parenthetical denotes the number of paddles of each length or the number of panels of each area.}
\end{table*}

Mechanical integration of the paddles is optimized for low mass, particularly near the scintillator itself, while maintaining the required robustness. 
Paddles are integrated within 21 carbon fiber panels (summarized in \autoref{tab:1pp} and illustrated in \autoref{fig:TOF_paddle_panel}), with small (\quant{1-3}{cm}) overlaps to minimize coverage gaps. 
The panels are fastened to the gondola using a system of custom spacers, which maintain a fixed offset while allowing for thermal expansion of the  Al gondola frame relative to the carbon fiber. 
Each preamp has three harnesses (LG signal, HG signal, and power) which are routed to minimize passive mass near sensitive detector elements. 

\begin{figure*}[tb]
\centering
\includegraphics[width=0.9\textwidth]{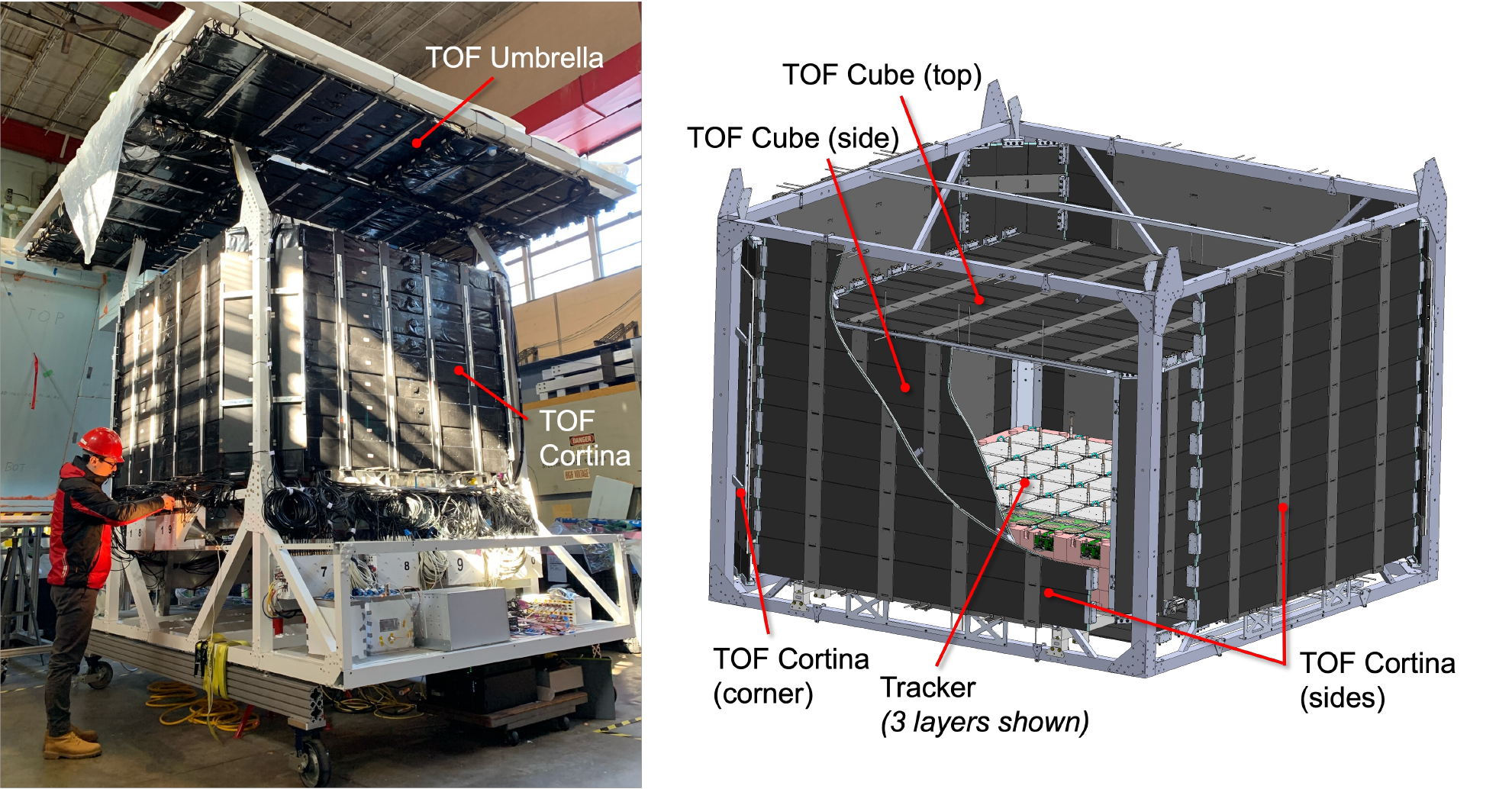}
\caption{\label{fig:cbe-cor-int}
In the fully integrated TOF ({\it left}), the Umbrella and Cortina are visible (graduate student for scale). The TOF Cortina, which encircles the Cube, is constructed from four side panels and four corner panels. The TOF Cube encloses the Tracker with 98\% hermeticity. For clarity, the engineering drawing ({\it right}) includes cutaways of the front Cube and Cortina panels and only three layers of the Tracker. 
}
\end{figure*}

The Cube, illustrated in \autoref{fig:cbe-cor-int}, is mounted directly onto the Tracker frame. 
Side panels are installed with horizontal paddles, \quant{4}{cm} 
from the Tracker on the radiator and boom faces and \quant{5}{cm} 
on the Sun and anti-boom faces, where the extra spacing accommodates Tracker harnesses and interface boards. 
The top (bottom) panel is mounted \quant{11}{cm} from the top Tracker layer (\quant{18}{cm} below the lowest layer), with the paddles oriented in the boom-antiboom (Sun-radiator) direction. 
To ensure hermiticity, both top and bottom panels extend laterally beyond the sides, and four vertical standalone paddles are mounted to the corners of the Tracker frame inboard of the side panels. 

The Cortina is mounted to the gondola Mid-frame, surrounding the Cube sides to enable time-of-flight measurements over a wide angular acceptance. As illustrated in \autoref{fig:cbe-cor-int}, the side panels are installed \quant{34}{cm} from the Cube sides with horizontal paddles.  
Four corner panels are installed at 45\degree\ to the side panels with paddles oriented vertically. 

The seven Umbrella panels are supported by the gondola Top-frame. 
The center panel is positioned \quant{93}{cm} above the Cube top. 
Six edge panels surround the central panel. These are mounted slightly higher than the central panel to allow for small overlaps that minimize coverage gaps at each adjoining panel edge.

\subsection{TOF Electronics}\label{sec:tofelec}

\subsubsection{Readout and Trigger Units}\label{sec:tofrat}

Twenty readout and trigger (RAT) units distributed throughout the gondola each contain a power board (PB), a local trigger board (LTB; see \autoref{sec:trigscheme}), and two readout boards (RBs; see \autoref{sec:rb}) necessary for the readout and trigger of eight paddles. Each PB provides power to sixteen preamps and all electronics in its RAT. 
Each output voltage (\quant{+3.3}{V}, \quant{+3.5}{V}, \quant{-1.5}{V}, and an adjustable SiPM bias of \quant{0-62}) is provided via isolated DC/DC converters. The PB and LTB are each controlled via Inter-Integrated Circuit (I2C) by a RB.

Distribution of the RAT units is optimized so all SiPM harnesses are \quant{\leq375}{cm} long while minimizing passive mass near sensitive detector elements.
Eleven RATs are located in the electronics bay, three RATS are installed along the lower perimeter of the Mid-frame, and the six RATs that service the Umbrella are installed in the Top-frame.

\subsubsection{Trigger Generation}\label{sec:trigscheme}

\begin{figure}[h]
\centering
\includegraphics[width=\columnwidth]{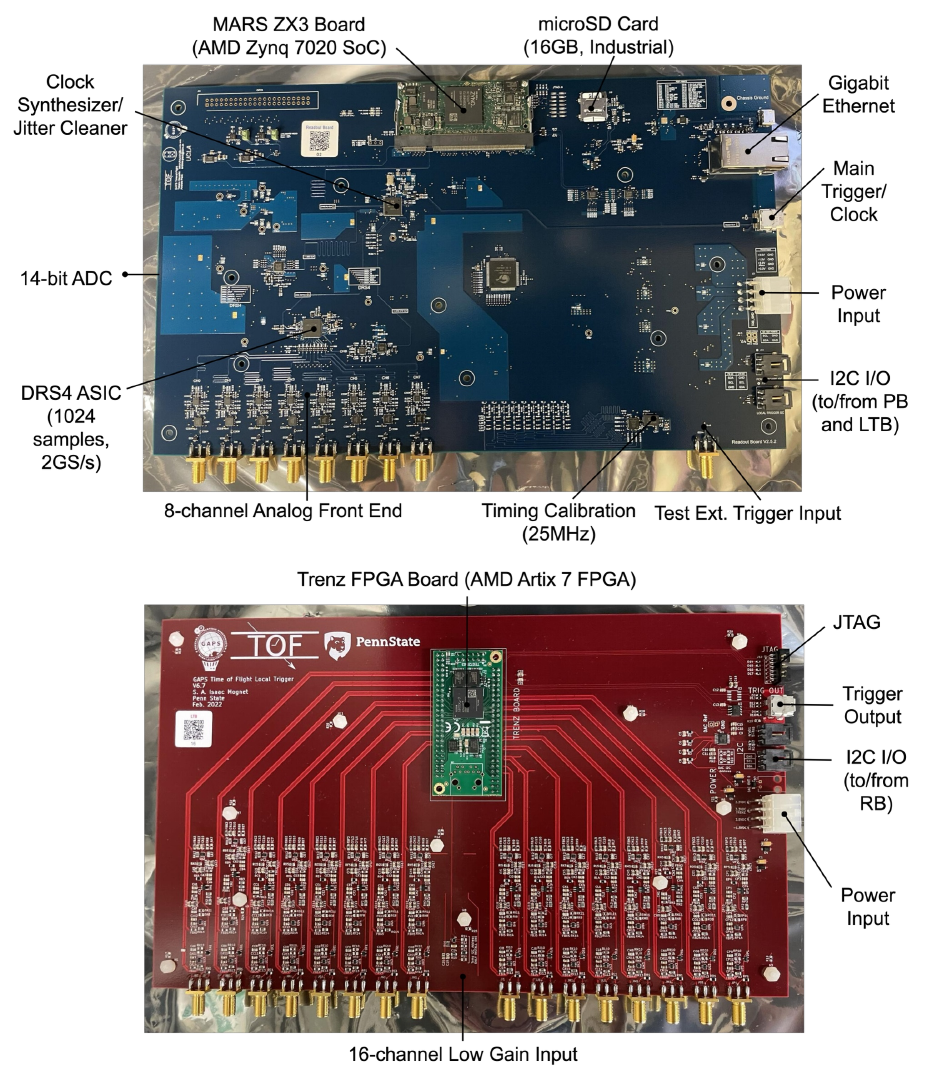}\caption{
TOF readout board (top) and local trigger board (bottom) with some of the key components identified. 
}
\label{fig:TOFelectronics}
\end{figure}

As illustrated in \autoref{fig:electronics}, trigger generation begins with the 20 independent LTBs. Each LTB (pictured in \autoref{fig:TOFelectronics}) transforms analog LG signals for each of its 16 channels (two ends of eight paddles) into a hit status for each paddle.
Each channel has three adjustable discriminators, as detailed in \autoref{tab:ltbdisc}, and the outputs from these discriminators are routed to the Trenz TE0725 FPGA board, which carries out the local trigger decision.
When any channel satisfies one of these thresholds, the LTB waits for input from the other channels during a \quant{25}{ns} coincidence window. 
Then, during a \quant{45}{ns} deadtime, it defines the hit status for each paddle as the highest discriminator level (\texttt{HIT}, \texttt{BETA}, \texttt{VETO}, or \texttt{NONE}) met by either end of the paddle and transmits the hit status for all eight paddles to the master trigger board (MTB) via Harting IE Cat.6A cables as a serial bit stream clocked at \quant{200}{Mbps}.

\begin{table}[ht]
\caption{\label{tab:ltbdisc} Discriminator levels in the LTB.}
\begin{tblr}{colspec={|lll|},hlines,width=\textwidth}
\centering
{Name}  & {Threshold$^*$} & {Usage}  \\
\hline
\texttt{HIT
} & \quant{40}{mV}$^{\dagger}$ & all charged particles\\
\texttt{BETA} & \quant{32}{mV}  & slow, highly ionizing particles\\ 
\texttt{VETO} & \quant{375}{mV}  & carbon \& higher-charge nuclei\\ 
\end{tblr}
\newline
\small
\textit{\raggedright *Thresholds are set relative to the most probable LG pulse height for a minimum ionizing particle (MIP) traversing 0.635\,cm of PVT, {\til12.5\,mV based on calibrations with atmospheric muons in a {180\,cm-long paddle.}}}\\
\textit{$^{\dagger}$To improve discrimination for small signals, the LG signal is passed through an op-amp with gain $4.7\pm0.4$ prior to the \texttt{HIT} discriminator circuit. Thus, the 40\,mV threshold can be compared to a most probable MIP pulse height of $\til 60$\,mV.}
\end{table}

The MTB combines information from all LTBs to generate a system trigger decision. 
The MTB is comprised of a Callisto Carrier Backplane that incorporates the Kintex 7 USB 3.1 FPGA module from Numato Lab and five DAQ Stack Interface daughter cards. 
The MTB deserializes the LTB signals to fill one hit status register for each paddle. The register persists for a configurable number of clock cycles to account for differences in the signal arrival for different LTBs. 
 To reduce systematic skew between paddle hits, the MTB also provides configurable delays for each LTB in increments of \quant{5}{ns}. These delays can be used to correct for fixed timing differences due to cable lengths.
The MTB implements an effective coincidence window by persisting the paddle registers for \quant{80}{ns}, after which they return to the \texttt{NONE} state.
After a system trigger (described below) is asserted, an MTB packet, including the paddle hit information associated with the trigger and the EVENT\_ID, is sent to the TOF CPU. Meanwhile, the MTB state is held for \quant{320}{ns}  (configurable in flight), during which no new triggers can be asserted. Paddle hit information received from any LTB during the holding window is added to the existing system trigger. Then, the status is reevaluated based on the \texttt{BUSY} signal from the Tracker. 
Considering the signal propagation time, LTB coincidence window, MTB coincidence window, and deadtime, the total time from a particle interaction in the TOF to the generation of a system trigger is \quant{\lesssim{400}}{ns}. The MTB time stamps system triggers using an external GPS pulse-per-second timestamp together with a \quant{10}{ns} local oscillator. 

Several trigger conditions are defined in the MTB firmware based on combinations of paddle hit status: 

\begin{enumerate}
\item{\texttt{Any}:} ${\geq}1$ paddle
\item{ \texttt{Track}:} ${\geq}1$ paddle (Cube) $+$ ${\geq}1$ paddle (Cortina or Umbrella)
\item{ \texttt{Track Central}:}  ${\geq}1$ paddle (Cube top) $+$ ${\geq}1$ paddle (Umbrella)
\item{ \texttt{Track Umbrella Central}:} ${\geq}1$ paddle (Umbrella center) $+$ ${\geq}1$ paddle (Cube top)
\item{\texttt{Antiparticle}}: several variations with different $p$, $q$, and $r$ are defined to satisfy all of: 
\begin{itemize}
\item $\geq p$ paddles (total)
\item $\geq q$ paddles (Umbrella $+$ Cortina)
\item $\geq r$ paddles (Cube)
\item $\geq 1$ \texttt{BETA} paddle (Cube top + sides)
\item $\geq 1$ \texttt{BETA} paddle (Umbrella + Cortina)
\end{itemize}
\end{enumerate}

The system trigger can be defined as any combination of the above trigger conditions, and each trigger condition can be individually disabled or prescaled.  
We note that the trigger conditions overlap (e.g.\ \texttt{Track Central} is a subset of \texttt{Track}). 

A handshake protocol links the MTB to the Tracker backend data acquisition:
\begin{enumerate}
\item The MTB monitors the status of the \texttt{BUSY} signal from the Tracker and only forms a system trigger if the Tracker is not \texttt{BUSY}.
\item When a system trigger forms, the MTB asserts a \texttt{TRIGGER} signal output to the Tracker. The \texttt{TRIGGER} signal remains until \texttt{BUSY} is received from the Tracker, indicating that the Tracker is processing the \texttt{TRIGGER}.
\item Once \texttt{BUSY} is received, the MTB distributes the \texttt{EVENT\_ID} to the Tracker on a \quant{1}{Mbps} LVDS link. The MTB also sends a packet indicating the triggered paddles and the \texttt{EVENT\_ID} directly to the relevant RBs.
\item The MTB re-enables itself to allow system triggers when the Tracker BUSY is de-asserted. 
\end{enumerate}

\subsubsection{Waveform Digitization}\label{sec:rb}

The RB (pictured in \autoref{fig:TOFelectronics}) digitizes HG waveforms following a trigger. There are 40 RBs, each with eight channels corresponding to the two ends of four paddles. 

The RB is built around the Enclustra Mars ZX3 board  which enables a Linux OS to coexist with an FPGA. 
The FPGA controls the digitization of TOF waveforms and transfers them to the OS memory. 
Data readout is performed by a Domino Ring Sampler 4 (DRS4) chip from Paul Scherrer Institute. The DRS4 is a switched capacitor array (SCA) with nine input channels and 1024 sampling cells per channel, operated at \quant{2}{GSPS}. One channel of each RB digitizes a \quant{20}{MHz} sinusoidal input supplied by the MTB for inter-RB timing calibration. The DRS4 produces one multiplexed analog output, which is converted to a digital signal by a 14-bit ADC chip. The digital signal is calibrated in both voltage and time (\autoref{sec:tofcal}). 

The RB digitizes waveforms upon receipt of an external (from the MTB) or internal \texttt{TRIGGER} signal. At this time, the FPGA halts sampling of the DRS4 chip, freezing the analog waveforms in the SCA. 
The DRS4 then reads the 1024 samples for each channel that is identified by the MTB as participating in the system trigger.
The resulting waveforms are packaged with the \texttt{EVENT\_ID} and metadata from the MTB and transferred from the FPGA to the OS memory using Direct Memory Access (DMA). 

OS memory is managed to balance efficiency and performance given high trigger rates for individual RBs. 
The OS memory reserved for the FPGA DMA is divided into two \quant{64}{MB} regions forming a Ping Pong buffer: at any time, half of the buffer is write-only and the other half is read-only. 
This mechanism allows the FPGA to continuously write to memory and enables the system to cope with high burst rates. 
Including metadata, each DRS4 waveform is \quant{2048}{B}, so that a full data packet including the sinusoid is \quant{\le18.5}{kB}. 
In ground calibration, individual RB rates are in the \squant{1-100}{Hz} range, leading to a high data rate overall.
As latency is crucial for the downstream event builder (\autoref{sec:tofevtbuild}), the read-out interval of the event buffer has been optimized to balance efficiency and performance. Several automatic schemes have been developed, and the software is adjustable at runtime through a control server running on the TOF CPU to account for different scenarios in flight. 

\subsubsection{TOF Event Builder}\label{sec:tofevtbuild}

TOF online activities are controlled by a multithreaded 
Rust-based software stack running on the TOF CPU. 
The TOF CPU is a single-board computer featuring a \quant{2.6}{GHz} Dual-Core
processor with \quant{16}{GB} of RAM and \quant{16}{TB} of storage. 
Both the TOF CPU and the MTB communicate through Gigabit Ethernet, which provides A32/D32 register access and burst access to data packets.

The primary purpose of the TOF online software is the assembly of RB data packets into a TOF event structure.
Data packets from the RBs arrive at the TOF CPU out of sequence, with latency depending on the hit multiplicity of the event and individual RB rates. 
The TOF online software calculates higher-level data products, such as hit energy and time-of-flight, from the waveforms in the RB packets. 
By matching \texttt{EVENT\_ID}s, the TOF software assembles information from the MTB packets containing the trigger information (including a list of paddles participating in the trigger) and the RB data packets into a TOF Event structure. After either all RB data packets {corresponding to the MTB trigger} are included or a configurable timeout window is met, the TOF software packages the MTB packet with summary information calculated from the RB packets and transmits it to the GCU. 
Any RB packets arriving after an event has timed out cannot be matched with an MTB packet based on \texttt{EVENT\_ID} and are thus excluded from the TOF Event. 

The TOF online software also controls the programmable trigger, waveform readout, and TOF event builder setting, and it manages housekeeping data collection, incoming command requests from the GCU, and transmission of events to the GCU. 
This software packages event and housekeeping data into a binary format that is passed to the GCU via Ethernet and separately saved on the TOF CPU. 
A single configuration file is used to update online settings including the trigger mode. 

\subsection{TOF Calibration and Operation}

\subsubsection{Calibration}\label{sec:tofcal}

The full set of TOF calibration data includes a RB DRS4 calibration for the waveforms, an overall gain calibration, a trigger threshold calibration, and a trigger prescale characterization.

The RB DRS4 waveform calibration uses internal RB inputs to characterize and correct the DRS4 effects on the digitized RB waveforms. 
{The 1024 sampling cells of each channel of the DRS4 operate in a ring mode, such that the starting point for the readout (starting cell) moves around in the ring from event to event. }
Each of the cells has a unique offset, gain, and timing width. Additionally, a time-dependent droop must be corrected for each waveform regardless of the starting cell. 
One thousand random triggers are generated with different starting cells using a \quant{0}{V} input and then using a \quant{182}{mV} input. 
The ADC offset is the mean value over the \quant{0}{V} triggers for a given cell, while the ADC gain is the difference between the mean value of the \quant{182}{mV} triggers and the \quant{0}{V} triggers for a given cell. 
The time-dependent voltage droop is corrected for by using the mean value for a given cell position relative to the starting cell. 
Finally, the timing width of each cell is calculated using the zero-crossing method detailed in \cite{Stricker_Shaver_2014} with 1000 random triggers on a \quant{25.0}{MHz} sine wave input.

The gain calibration adjusts the SiPM bias voltages to ensure that the TOF paddles are uniformly sensitive to minimum ionizing particles and can be compared for precision timing and position reconstruction. 
Data are collected using the \texttt{Track} trigger condition, delivering a sample dominated by MIP events. Starting at a nominal bias voltage of \quant{+58}{V} on each paddle, voltages on all paddles are adjusted  by up to \quant{\pm1.0}{V} to obtain an average charge per event of \quant{15-30}{pC} on the horizontal panels (Umbrella, Cube top, and Cube bottom), where down-going particles are largely normally incident. 
Following this first inter-paddle adjustment, the gain is balanced between two ends of the same paddle. The distribution of timing difference between the two paddle ends systematically skews away from the lower-gain SiPM, so the gain is balanced by adjusting the relative voltages until the timing distribution on a paddle is balanced. 
The Cube side and Cortina panels record higher average charges due to the larger incident angles and therefore are not gain-adjusted. 
The SiPM gain is temperature dependent and thus different in flight compared to the ground. The bias voltages can be adjusted in flight to account for temperature changes relative to ground calibrations, using the temperature dependence of $0.054\,\textrm{V/C}\degree$ as measured in the laboratory.

The threshold behavior of the LG trigger inputs are characterized following gain adjustments. Data were collected varying the LTB thresholds for all paddles in unison. On the ground, the parameter space around the nominal \quant{40}{mV} \texttt{HIT} threshold is sampled using data are collected using the \texttt{Track} trigger (\squant{500}{Hz}), with the \texttt{HIT} threshold set to 20, 30, 40, 50, 60, 80, 100, and \quant{120}{mV} (downstream of the op amp, \autoref{tab:ltbdisc}). 
A \texttt{BETA} threshold study can be performed in flight {by making use of} the higher energy depositions from slower particles that do not reach the ground, using 
\texttt{Antiparticle} trigger 
with the nominal \texttt{HIT} threshold and varying the \texttt{BETA} threshold. 

The prescale parameter permits the system trigger to randomly accept a fraction of the events satisfying a given hardware trigger. Precise characterization of this parameter, based on comparing the trigger rates resulting from data collection with a range of prescale values in the absence of dead time, enabled calculation of the absolute trigger rates. 

\begin{figure}[tb]
\centering
\includegraphics[width=0.9\columnwidth,trim = 5 5 20 50,clip]{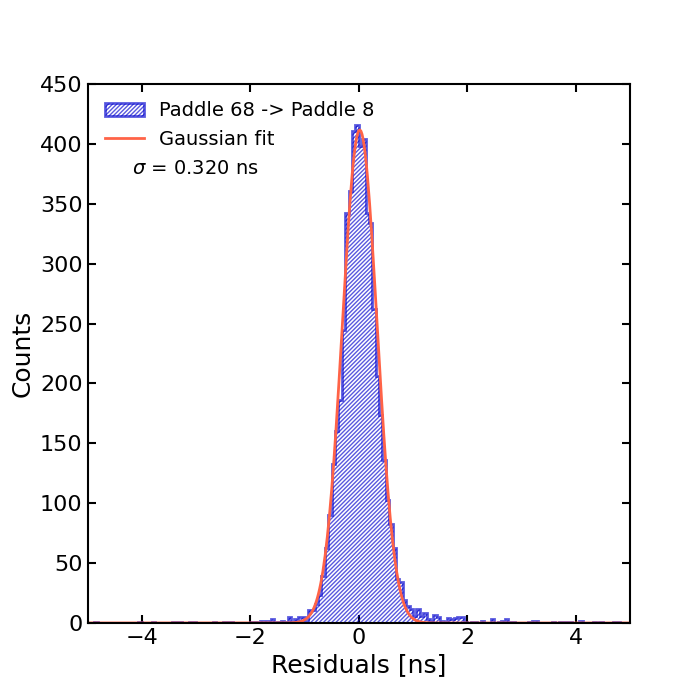}
\caption{\label{fig:tof_timing}
Distribution of time-of-flight residuals between paddle 68 (in the Umbrella) and paddle 8 (directly below 68, in the top of the Cube) during ground commissioning with cosmic-ray muons. A Gaussian fit indicates a timing resolution of \squant{0.320}{ns}, well under the \quant{0.400}{ns} requirement.
}
\end{figure}

Following calibration of the system, the timing resolution performance was demonstrated using cosmic-ray muons. \autoref{fig:tof_timing} illustrates the distribution of residuals relative to the true time-of-flight for a paddle traversing two specific paddles at the speed of light. The fitted time-of-flight resolution of \quant{0.320}{ns} (1$\sigma$) is smaller than the \quant{0.400}{ns} requirement.

\subsubsection{Operation}

The nominal operation mode has a system trigger that is a combination of two or more trigger conditions, each operated with its own prescale. 
On the ground, the \texttt{Antiparticle} conditions are only triggered at a rate of a few Hz, and most of the events in typical calibrations were obtained using the \texttt{Track} trigger.

\section{Trigger interface and full system readout}\label{sec:sys}

\autoref{fig:electronics} diagrams the trigger and readout connections of the GAPS gondola electronics, as well as the system-level Tracker and TOF electronics. 

\subsection{Trigger Interface and GPS Timecode}\label{sec:tiu}
A dedicated trigger interface unit (TIU) facilitates low-latency direct communication between Tracker and TOF. The TIU
distributes the TOF \texttt{TRIGGER} and \texttt{EVENT\_ID} signals to the Tracker's backend data acquisition units and returns their \texttt{BUSY} signal to the MTB. 

Additionally, the TIU distributes a GPS timestamp to the TOF and the Tracker once per second, enabling an absolute timestamp for triggered events and providing an independent method to check the correct association of Tracker and TOF data. The NASA/CSBF Trimble GPS unit provides the timestamp and pulse-per-second (PPS) signal to a custom timecode generator unit, which in turn provides a synchronized timecode to the TIU.

\subsubsection{GAPS Event Builder}\label{sec:intevts}
The online GCU software merges the science data from the Tracker and TOF subsystems by matching TOF packets and Tracker packets by \texttt{EVENT\_ID}. When the GCU receives the first packet with a new \texttt{EVENT\_ID}, it then waits for additional packets with the same \texttt{EVENT\_ID} arriving during an adjustable time period (nominally {\quant{120}{s}}). Then, {all of} the Tracker and TOF packets {with the same \texttt{EVENT\_ID}} are packaged into a merged event packet to be written to disk. Any Tracker or TOF packet {belonging to a given triggered event that arrives} after this period causes a second merged event packet with the same \texttt{EVENT\_ID}. 

Considering the in-flight bandwidth of the various telemetry systems described in \autoref{sec:tmcm}, only a portion of the science data can typically be transmitted to the ground for real-time data-quality monitoring and science analysis. The GCU therefore performs an online classification to flag antinucleus-like events for priority transmission. 

The online classification is based on six variables: the multiplicity and total energy deposition of Tracker hits above a programmable software energy threshold, the multiplicity and total energy deposition of hits in the TOF Cube, and the multiplicity and total energy deposition of hits in the outer TOF (Umbrella + Cortina). To be prioritized for telemetry, an event must be within an allowed range on each of these variables. The allowed ranges are adjustable in flight, enabling a less restrictive classification if the telemetry bandwidth is higher. 

In addition to prioritizing antinucleus-like events for telemetry, the GCU further controls the telemetry bandwidth by imposing a programmable software energy threshold on the Tracker data within an event. Below the threshold, {hits are} saved to disk but not included in the telemetry stream.

\section{Conclusion}\label{sec:conclusion}

GAPS is an Antarctic LDB mission specifically designed to meet the challenges of low-energy cosmic-ray antinucleus detection. 
The  payload is the first instrument capable of enabling cosmic-ray antinucleus science spanning the kinetic energy range of \quant{0.10-0.25}{GeV/{\it n}}, including sensitivity to the cosmic-ray antideuteron fluxes predicted from dark matter. The Tracker provides a high-acceptance target that can stop cosmic-ray particles in this energy range while resolving the exotic atom de-excitation X-rays and annihilation products. The TOF provides a high-speed system trigger with a wide-angle acceptance and a velocity measurement with $\Delta \beta  (1\sigma) < 0.1c$, as well as additional tracking information. The GAPS sensitivity is possible on a balloon platform thanks to the novel exotic atom-based particle identification method that delivers robust background rejection without a magnet.

The first GAPS science payload was integrated, tested, and calibrated on the ground from 2022 -- 2024, with the final integration at the Long-duration Balloon facility at McMurdo Station, Antarctica during NASA's 2024/25 Antarctic balloon campaign. Because weather conditions precluded launch during the 2024/25 campaign, the payload flew during NASA's 2025/26 Antarctic balloon campaign. This publication details the sensitive detector systems and engineering design of the GAPS Antarctic balloon payload. 

\section*{Acknowledgment}

We are grateful to the team at Columbia Scientific Balloon Facility for their enduring commitment to letting GAPS fly and for their superlative  support of our team and our science over the past several years. 
We thank NASA's Balloon Program Office for their support. 
We are also indebted to the Antarctic Support Contract's LDB team who moved mountains (of snow and of snacks) to keep the balloon facility running through all conditions.

Sincere thanks are due to the staff at MIT's Bates Laboratory, the University of California, Berkeley's Space Sciences Laboratory (SSL), and Columbia University's Nevis Laboratory for their support during the on-site integration and testing of the GAPS science payload. 

The GAPS team at SSL is grateful to the NCT, COSI, and GRIPS balloon teams for a tradition of sharing resources and expertise to allow mutual success. We also thank John Tomsick for his continued support of the GAPS program and specifically for enabling the contributions of F.\ Rogers.

This work is supported in the U.S. by NASA Astrophysics Research and Analysis GAPS grants
(NNX17AB44G, NNX17AB46G, NNX17AB47G, 80NSSC21K1877) PI, C. Hailey, in Japan by
JAXA/ISAS Small Science Program FY2017, and in Italy by Istituto Nazionale di Fisica Nucleare
(INFN) and by the Italian Space Agency through the ASI INFN agreement n.\  2018-28-HH.0:\ 
``Partecipazione italiana al GAPS - General AntiParticle Spectrometer''. H. Fuke is supported by
JSPS KAKENHI grants (JP17H01136, JP19H05198, JP22KK0042, and JP22H00147). K.\ Perez, G.\ Bridges, K.\ Pappas, and S.\ Vickers are  supported by Heising-Simons Foundation awards 2023-4617 and 2025-6055. R.A.\ Ong receives support from the UCLA Division of Physical Sciences.
Oak Ridge National  Laboratory is operated by UT-Battelle, LLC, under contract DE-AC05-00OR22725 with the U.S. Department of Energy (DOE). S.\ Feldman was supported through the National Science Foundation Graduate Research Fellowship under grant 2034835. The contributions of C.\ Gerrity were supported by NASA under award No. 80NSSC19K1425 of the Future Investigators in NASA Earth and Space Science and Technology (FINESST) program. K.\ Yee is supported through the National Science Foundation Graduate Research Fellowship under grant 2141064. K.\ Aoyama receives support from JSPS KAKENHI grant JP24K22891. M.\ Kozai receives support from JSPS KAKENHI grant JP20K14505. K.\ Mizukoshi receives support from JSPS KAKENHI grants (JP22K20375 and JP24K17079). S.\ Okazaki receives support from JSPS KAKENHI grants (JP18K13928 and JP22KK0042). Y.\ Shimizu receives support from JSPS KAKENHI grants (JP20K04002, JP22KK0042, JP23K03436). M.\ Xiao, Z.\ Wu and J.\ Yang receive support from Yangyang Development Fund. 

The technical support and advanced computing resources from University of Hawaii Information Technology Services – Cyberinfrastructure, funded in part by the National Science Foundation MRI Grant No.\ 1920304, are appreciated. This research was performed using resources provided by the Open Science Grid \cite{osg07,osg09,OSGPool,OSDF}, which is supported by the National Science Foundation Grant Nos.\ 2030508 and 2323298

\bibliographystyle{elsarticle-num}
\bibliography{main}

\end{document}